\documentclass[
reprint,
superscriptaddress,
amsmath,amssymb,
aps,
]{revtex4-2}

\usepackage{color}
\usepackage{graphicx}% Include figure files
\usepackage{dcolumn}% Align table columns on decimal point
\usepackage{bm,braket}% bold math
\begin {document}
%section {title}
%\preprint{APS/123-QED}

\title{
    Symmetry breaking of current response in disordered exclusion processes
    }% Force line breaks with \\

\author{Issei Sakai}
\email{6224701@ed.tus.ac.jp}
\affiliation{%
    Department of Physics and Astronomy, Tokyo University of Science, Noda, Chiba 278-8510, Japan
}%

\author{Takuma Akimoto}
\email{takuma@rs.tus.ac.jp}
\affiliation{%
    Department of Physics and Astronomy, Tokyo University of Science, Noda, Chiba 278-8510, Japan
}%

%}


\date{\today}% It is always \today, today,

\begin{abstract}
    The bias-reversal symmetry---where reversing an external bias inverts the current without changing its magnitude---is a hallmark of nonequilibrium transport.
    While this property holds in homogeneous systems such as the asymmetric simple exclusion process, how disorder and its interplay with particle interactions affect this symmetry has remained unclear.
    Here, we identify a general criterion in disordered exclusion processes showing that the bias-reversal symmetry holds if and only if the local left-right bond-bias ratio is spatially uniform.
    This criterion provides a practical diagnostic that separates heterogeneous environments into symmetry-preserving and symmetry-breaking classes.
    Mean-field and numerical analyses reveal that bond disorder preserves the symmetry beyond linear response, whereas site disorder breaks it through an interplay between heterogeneity and particle interactions.
    Our results demonstrate how environmental disorder and interparticle interactions cooperate to generate asymmetric transport, thereby providing insight that is potentially relevant to transport through biological and artificial nanochannels.
\end{abstract}

%\pacs{05.45.Ac, 05.40.Fb, 87.15.Vv}% PACS, the Physics and Astronomy
\maketitle

%\tableofcontents

\section{Introduction}
The bias-reversal symmetry---where reversing the applied bias inverts the sign of a transport observable while preserving its magnitude---is a fundamental property of nonequilibrium transport.
It is rigorously preserved in homogeneous systems, such as the asymmetric simple exclusion process (ASEP) \cite{SPITZER1970246,DERRIDA199865,Derrida_2007}, in which particles driven by an external field diffuse on a one-dimensional lattice without overtaking each other, even beyond the linear-response regime~\cite{De-Masi:1985aa,Derrida:1997aa,4j5q-j4ht}.
Yet, in realistic heterogeneous environments, this symmetry may either persist or break, and how disorder and interparticle interactions jointly determine its fate remains unclear.

Understanding when and why the bias-reversal symmetry breaks is essential for describing transport in heterogeneous media, such as biological ion channels \cite{Misakian:2003aa,NESTOROVICH20033718,NOSKOV20042299,AKSIMENTIEV20053745,10.1085/200609655,Bhattacharya:2011aa,Piguet:2014aa,PAYET20151600,Manara:2015aa,https://doi.org/10.1002/cbic.201600644,Zhou:2020aa,Dessaux:2022aa,membranes13050517} and artificial nanopores \cite{https://doi.org/10.1002/adfm.200500471,10.1063/1.4776216,Gamble:2014aa,SU2023141064}, where structural or energetic asymmetry is coupled to particle interactions.
In such systems, disorder can act either as a passive perturbation or as an active source of rectification, depending on how it modifies local transition rates.
Despite extensive studies of disordered exclusion processes~\cite{PhysRevLett.78.3039,PhysRevE.58.1911,PhysRevE.58.4226,Enaud:2004aa,PhysRevE.70.016108,PhysRevLett.94.010601,PhysRevE.74.061101,PhysRevE.75.011127,Greulich:2008aa,PhysRevE.78.061116,PhysRevLett.112.050603,10.1214/14-BJPS277,PhysRevResearch.2.013025,PhysRevResearch.2.043073,Goswami:2022aa,PhysRevE.107.L052103,PhysRevE.107.054131}, a general criterion linking environmental heterogeneity to the symmetry of the current response has remained elusive.

In homogeneous systems, such as the standard ASEP, the stationary current satisfies $J(\rho,\varepsilon)=\varepsilon\rho(1-\rho)$~\cite{Derrida:1997aa}, which exactly obeys both the bias-reversal symmetry, $J(\rho,\varepsilon)=-J(\rho,-\varepsilon)$, and the particle-hole symmetry, $J(\rho,\varepsilon)=J(1-\rho,\varepsilon)$.
These relations reflect a simple correspondence between particle and hole dynamics under the same external bias.
However, once spatial heterogeneity is introduced---through site- or bond-dependent hopping rates---this correspondence does not necessarily hold, and the conditions under which the symmetries persist or break remain unclear.

To address this open question, we investigate the particle current in the ASEP under heterogeneous environments.
We derive a general condition showing that when the ratio of forward to backward hopping rates is uniform across all bonds, the bias-reversal and particle-hole symmetries hold simultaneously.
In contrast, when this ratio varies spatially, at least one of these symmetries must be broken.
Mean-field and numerical analyses of ASEPs on quenched random energy landscapes verify these statements: bond disorder preserves the symmetries beyond the linear-response regime, whereas site disorder breaks them through the cooperative interplay between heterogeneity and particle interactions.

\section{Model}
We consider a disordered ASEP.
The system consists of $N$ particles on a one-dimensional lattice with $L$ sites under periodic boundary conditions.
Because of the exclusion principle, each site can be occupied by at most one particle.
Particles perform biased random walks in a heterogeneous environment.
A particle at site $i$ attempts to hop to site $i-1$ with rate $q_i(\varepsilon)$ and to site $i+1$ with rate $p_i(\varepsilon)$, where
\begin{equation}
    q_i(\varepsilon)=\frac{1-\varepsilon}{2}w_{i,i-1},\quad
    p_i(\varepsilon)=\frac{1+\varepsilon}{2}w_{i,i+1}.
\end{equation}
Here, $\varepsilon\in[-1,1]$ is the bias parameter that controls the strength and direction of the external drive:
$\varepsilon=0$ corresponds to equilibrium dynamics, while $\varepsilon\neq0$ drives the system out of equilibrium. 
The coefficients $w_{i,i\pm1}/2$ represent the hopping rates from site $i$ to $i\pm1$ in the corresponding equilibrium system and encode the heterogeneity of the environment.
In the unbiased case ($\varepsilon=0$), the rates satisfy $\prod_{i=1}^Lw_{i,i-1}/w_{i,i+1}=1$ ensuring that the stationary current vanishes in equilibrium (see Appendix~\ref{appendix: A}).

We focus on the stationary current $J(\rho,\varepsilon)$, defined as the average number of particle hops across a bond per unit time for density $\rho$ and the bias $\varepsilon$:
\begin{equation}
    J(\rho,\varepsilon)=\lim_{t\rightarrow\infty}\frac{\Braket{Q_i(t)}}{t},
\end{equation}
where $\Braket{\cdot}$ denotes the ensemble average and $Q_i(t)$ is the integrated current between site $i$ and $i+1$ at time $t$.
In the steady state, the continuity equation ensures that the current is independent of the bond, i.e., $J(\rho,\varepsilon)$ does not depend on $i$.
We analyze how the stationary current depends on the bias and density.
In particular, we examine the bias-reversal symmetry
\begin{equation}
    J(\rho,\varepsilon)=-J(\rho,-\varepsilon)
\end{equation}
and the particle-hole symmetry
\begin{equation}
    J(\rho,\varepsilon)=J(1-\rho,\varepsilon).
\end{equation}

\begin{figure}[t]
    \centering
    \includegraphics[width=7cm]{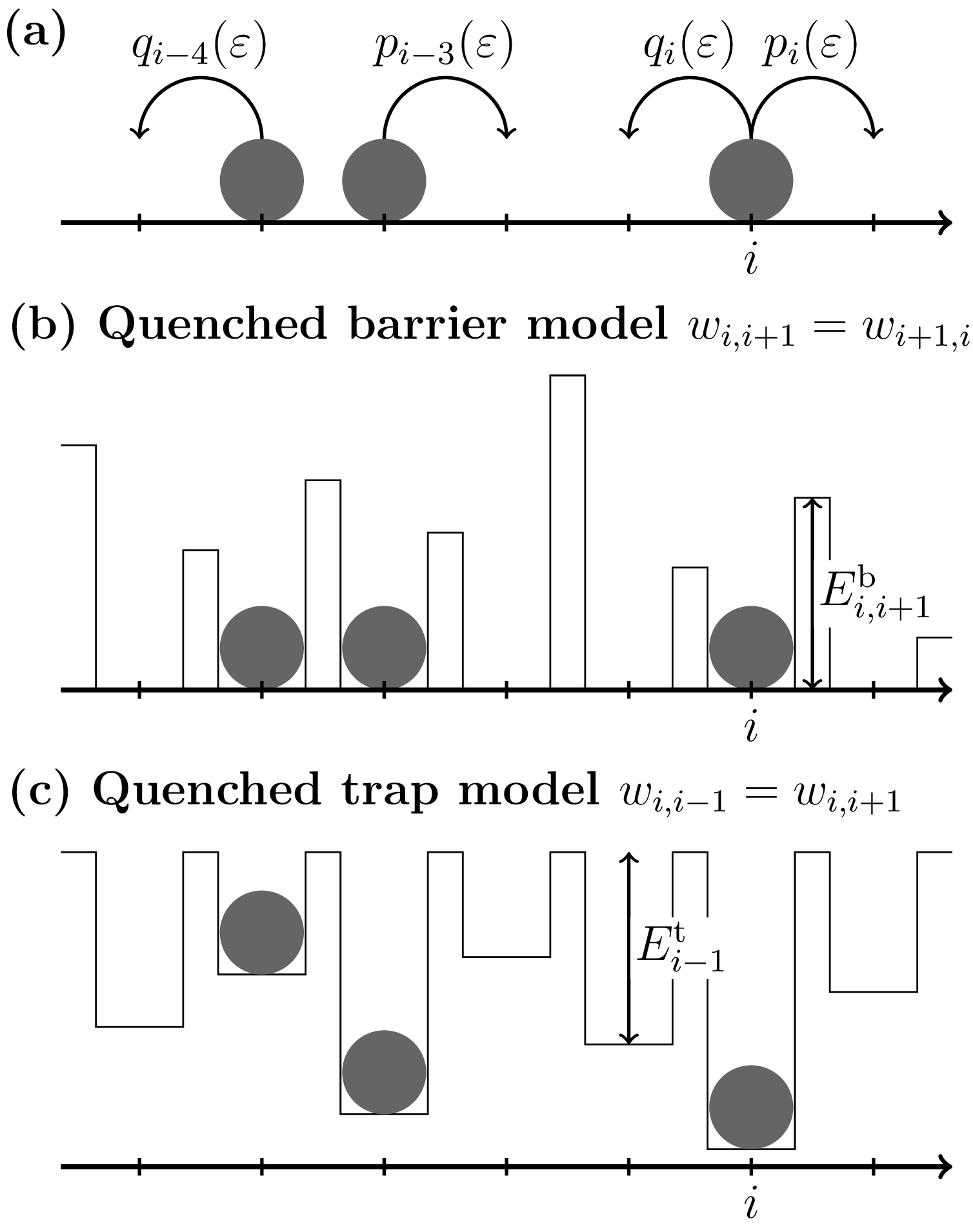}
    \caption{Schematic illustration of the disordered ASEP.
    (a) Particles hop on a one-dimensional lattice with site-dependent left/right rates.
    (b) Quenched barrier model, in which each bond has a symmetric barrier.
    (c) Quenched trap model, in which each site has a symmetric trap.}
    \label{fig: model}
\end{figure}

To analyze these properties, we consider two concrete realizations of heterogeneous environments: (i) the quenched barrier model (QBM) and (ii) the quenched trap model (QTM) \cite{BouchaudGeorfes}.
Both models describe random walks in quenched random energy landscapes [Figs.~\ref{fig: model}(b) and (c)], where the transition rates obey an Arrhenius law, $w_{i,i\pm1}=\tau_c^{-1}\exp(-E_{i,i\pm1}/T)$, with $E_{i,i\pm1}$ the local energy barrier, $T$ the temperature, and $\tau_c$ a characteristic time.
The energy landscape is quenched, meaning that the random energy landscape remains fixed in time.
In the QBM, each bond has a symmetric barrier $E_{i,i+1}^{\mathrm{b}}$, i.e., $w_{i,i+1}=w_{i+1,i}$, which we refer to as bond-symmetric disorder.
In contrast, in the QTM, each site has a symmetric trap $E_{i}^{\mathrm{t}}$, which implies $w_{i,i-1}=w_{i,i+1}$ and corresponds to site-symmetric disorder.
The barrier heights $E_{i,i+1}^{\mathrm{b}}$ and trap depths $E_i^{\mathrm{t}}$ are independent and identically distributed random variables drawn from the exponential distribution 
$\phi(E)=T_g^{-1}\exp(-E/T_g)$, where $T_g$ denotes the glass temperature.

Notably, both the barrier and trap energies are locally left-right symmetric, implying that neither model contains any intrinsic microscopic bias.
One might therefore expect both environments to preserve the bias-reversal symmetry.
However, as we show below, site-symmetric disorder in the QTM can nevertheless break this symmetry through its interplay with particle interactions.

\section{Results}
\subsection{Criterion for Symmetric Transport in Disordered ASEP}
We define the bond-bias ratio
\begin{equation}
    r_i(\varepsilon)=\frac{p_i(\varepsilon)}{q_{i+1}(\varepsilon)},
\end{equation}
which quantifies the local left-right bias across bond $(i,i+1)$.
Our central result is that the bias-reversal symmetry $J(\rho,\varepsilon)=-J(\rho,-\varepsilon)$ and the particle-hole symmetry $J(\rho,\varepsilon)=J(1-\rho,\varepsilon)$ are equivalent for all densities if and only if $r_i(\varepsilon)$ is independent of $i$.

We first show that a uniform bond-bias ratio implies the equivalence of two symmetries.
When $r_i(\varepsilon)$ is uniform, the local left-right bias is spatially constant.
A hole at site $i$ moves left when a particle hops from $i-1$ to $i$ with rate $p_{i-1}(\varepsilon)$ and moves right when a particle hops from $i+1$ to $i$ with rate $q_{i+1}(\varepsilon)$.
Thus, the hole at bias $\varepsilon$ has left and right hopping rates $p_{i-1}(\varepsilon)$ and $q_{i+1}(\varepsilon)$, respectively.
A particle at site $i$ under the reversed bias $-\varepsilon$ hops left with rate $q_i(-\varepsilon)$ and right with rate $p_i(-\varepsilon)$.
For a uniform bond-bias ratio (in particular, under bond-symmetric connectivity $w_{i,i+1}=w_{i+1,i}$), these rates coincide, $p_{i-1}(\varepsilon)=q_i(-\varepsilon)$ and $q_{i+1}(\varepsilon)=p_i(-\varepsilon)$, so that a hole at bias $\varepsilon$ evolves in exactly the same way as a particle at bias $-\varepsilon$.
Denoting the hole current at hole density $\rho$ and bias $\varepsilon$ by $J_{\mathrm{hole}}(\rho,\varepsilon)$, this dynamic equivalence implies $J_{\mathrm{hole}}(\rho,\varepsilon)=J(\rho,-\varepsilon)$.
On the other hand, a hole current is simply the particle current viewed from empty sites, and therefore $J_{\mathrm{hole}}(\rho,\varepsilon)=-J(1-\rho,\varepsilon)$.
Combining these two relations yields
\begin{equation}
    J(1-\rho,\varepsilon)=-J(\rho,-\varepsilon).
    \label{eq: current relation for bond-bias symmetry}
\end{equation}
Equation~\eqref{eq: current relation for bond-bias symmetry} is structurally analogous to the symmetry-based current relation derived in the Appendix of Ref.~\cite{PhysRevE.58.1911}, although the underlying model is different.
Equation \eqref{eq: current relation for bond-bias symmetry} provides a direct link between the two symmetries.
If the particle-hole symmetry $J(\rho,\varepsilon)=J(1-\rho,\varepsilon)$ holds, then Eq.~\eqref{eq: current relation for bond-bias symmetry} immediately gives the bias-reversal symmetry $J(\rho,\varepsilon)=-J(\rho,-\varepsilon)$.
Conversely, if the bias-reversal symmetry holds, inserting $J(\rho,\varepsilon)=-J(\rho,-\varepsilon)$ into Eq.~\eqref{eq: current relation for bond-bias symmetry} yields $J(1-\rho,\varepsilon)=J(\rho,\varepsilon)$, i.e., the particle-hole symmetry.
Thus, under a uniform bond-bias ratio, the bias-reversal and particle-hole symmetries are equivalent for all densities.

Conversely, suppose that the bias-reversal and particle-hole symmetries are equivalent for all $\rho$.
Then Eq.~\eqref{eq: current relation for bond-bias symmetry} must hold, which means that the dynamics of holes at bias $\varepsilon$ can be mapped onto those of particle at bias $-\varepsilon$ for all densities.
Such a mapping is possible only when the local left-right bias is the same on every bond, i.e., when the bond-bias ratio $r_i(\varepsilon)$ is independent of $i$.
Therefore, the bias-reversal and particle-hole symmetries are equivalent for all densities if and only if the bond-bias ratio is uniform across the system.

\subsection{Disorder-induced symmetry breaking}
When the bond-bias ratio $r_i(\varepsilon)$ varies from bond to bond, the local left-right bias is no longer uniform, and the dynamics of holes under bias $\varepsilon$ differ from those of particles under the reversed bias $-\varepsilon$.
In this case, the bridge relation $J(1-\rho,\varepsilon)=-J(\rho,-\varepsilon)$ fails, so at least one of the two symmetries---the bias-reversal symmetry or the particle-hole symmetry---must be violated.
Consequently, the two symmetries cannot hold simultaneously, and the environmental heterogeneity can induce a breakdown of the bias-reversal symmetry, leading to an asymmetric current response.

\subsection{Linear-response regime}
In the linear-response regime ($|\varepsilon|\ll 1$), the equilibrium current fluctuations and the average current induced by a weak bias are related through
\begin{equation}
    J(\rho,\varepsilon)
    \sim \frac{\varepsilon}{L}\lim_{t\to\infty}\frac{d\Braket{Q(t)^2}_0}{dt}\quad(\varepsilon\rightarrow0),
    \label{eq: linear response relation}
\end{equation}
which is a form of fluctuation-dissipation relation: the linear-response coefficient is determined by the equilibrium fluctuations of the integrated current.
Importantly, this relation holds even in heterogeneous many-particle systems, including the disordered ASEP considered here.
Here, $\Braket{\cdot}_{0}$ denotes an equilibrium average, and $Q(t)$ is the total integrated current, defined as $Q(t)=\sum_{i=1}^LQ_i(t)$.
We derive Eq.~\eqref{eq: linear response relation} by following the argument of Ref.~\cite{Hanney:2003aa}, and a detailed derivation in our notation is provided in Appendix~\ref{appendix: B}.

Equation~\eqref{eq: linear response relation} implies that $J(\rho,\varepsilon)=-J(\rho,-\varepsilon)$ for $|\varepsilon|\ll1$,
so the current is an odd function of the bias parameter in the linear-response regime.
When the bond-bias ratio $r_i(\varepsilon)$ is independent of $i$, this bias-reversal symmetry, together with the criterion established above, ensures that the particle-hole symmetry $J(\rho,\varepsilon)=J(1-\rho,\varepsilon)$ also holds.
In contrast, if the ratio depends on $i$, the mapping between particles and holes under bias reversal breaks down and the particle-hole symmetry is violated, i.e., $J(\rho,\varepsilon)\neq J(1-\rho,\varepsilon)$.
For the QTM, the equilibrium current fluctuations exhibit an asymmetric dependence on the density~\cite{PhysRevE.111.014134}.

\subsection{Nonlinear-response regime}
Beyond the linear-response regime, the heterogeneous environment gives rise to nonlinear transport behavior. 
A key question is whether the next-order contribution beyond the linear term in $\varepsilon$ appears at order $\varepsilon^2$ or $\varepsilon^3$.
If the term arises at order $\varepsilon^2$, the current exhibits an asymmetric dependence on the sign of the bias, i.e., $J(\rho,\varepsilon)\neq-J(\rho,-\varepsilon)$, reflecting the asymmetric response.

We next identify the conditions under which the next-order contribution appears at order $\varepsilon^3$.
Two distinct types of conditions ensure this behavior: (i) the bond-independent ratio condition or (ii) the inversion-correspondence condition.
(i) The bond-independent ratio condition requires that the bond-bias ratio $r_i(\varepsilon)$ is independent of the bond index $i$.
As discussed above, this condition ensures the bias-reversal symmetry $J(\rho,\varepsilon)=-J(\rho,-\varepsilon)$.
(ii) The inversion-correspondence condition identifies cases in which the bias-reversal symmetry can still hold even when the bond-bias ratio is not uniform.
This inversion-correspondence yields the bias-reversal symmetry because a spatial inversion of the lattice, combined with reversing the bias, maps each microscopic hop at $\varepsilon$ onto a corresponding hop at $-\varepsilon$.
As a result, the forward- and backward-biased processes are dynamically equivalent, leading to $J(\rho,\varepsilon)=-J(\rho,-\varepsilon)$ (see Appendix~\ref{appendix: C} for the detailed proof).
At the same time, since the ratio $r_i(\varepsilon)$ depends on the bond, the particle-hole symmetry is broken.
For disordered systems, the inversion-correspondence condition is a rare event, since it requires a finely tuned spatial symmetry of the random rates.
We conjecture---and later confirm numerically---that an $\varepsilon^{2}$ contribution appears whenever neither condition (i) nor condition (ii) is satisfied.

\begin{figure}[b]
    \centering
    \includegraphics[width=8.2cm]{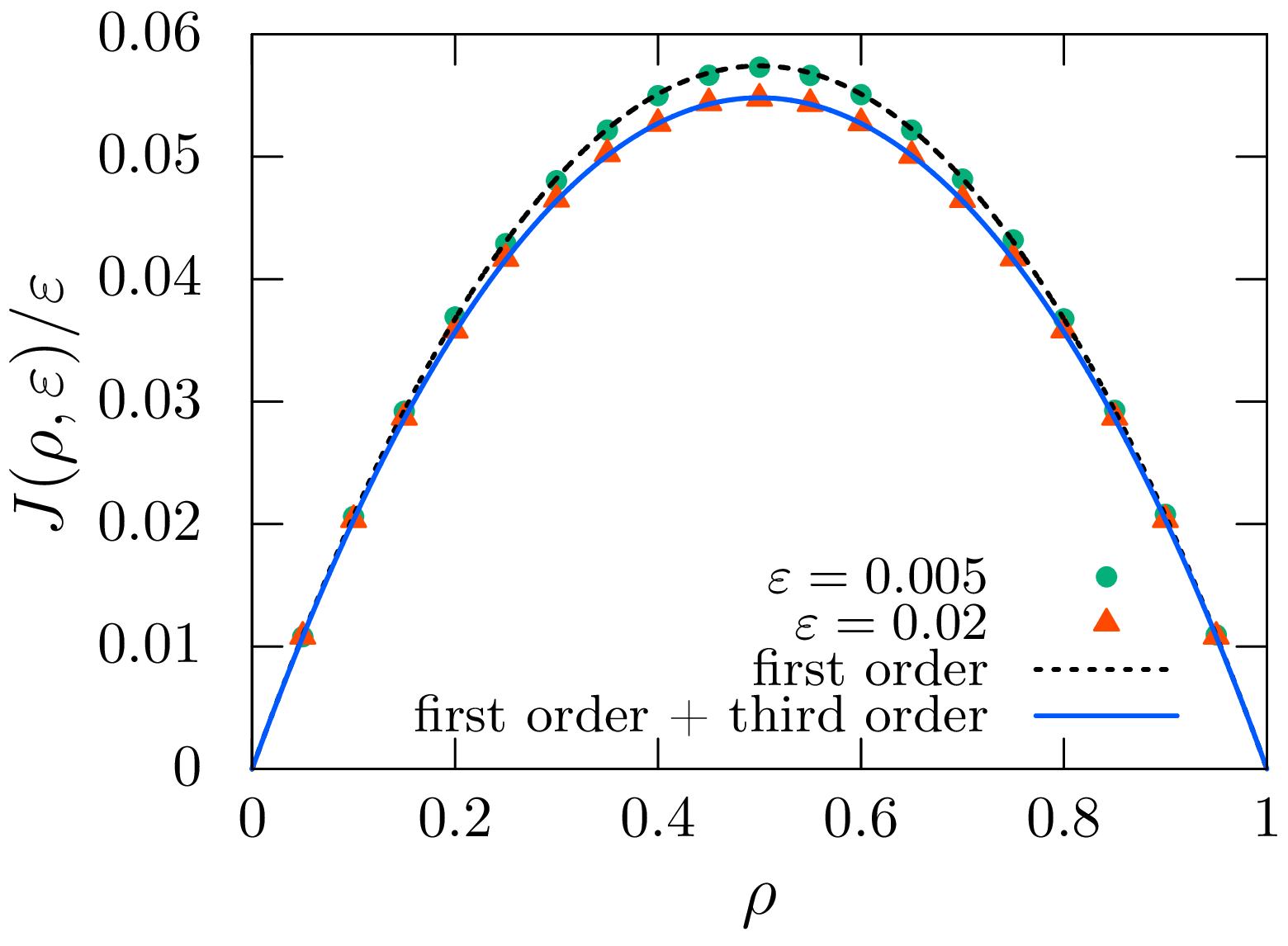}
    \caption{Current-density relation in the ASEP with the quenched barrier model ($L=100$, $T/T_g=1.5$, $\tau_c=1$).
    The vertical axis shows the scaled current $J(\rho,\varepsilon)/\varepsilon$.
    Symbols represent numerical results obtained for a single realization of the quenched random energy landscape (circles: $\varepsilon=0.005$; triangles: $\varepsilon=0.02$).
    Dashed lines correspond to the first-order term of Eq.~\eqref{eq: current in QBM}, and solid lines include the third-order correction for $\varepsilon=0.02$.
    }
    \label{fig: current-density}
\end{figure}

\begin{figure*}[t]
    \centering
    \includegraphics[width=17.2cm]{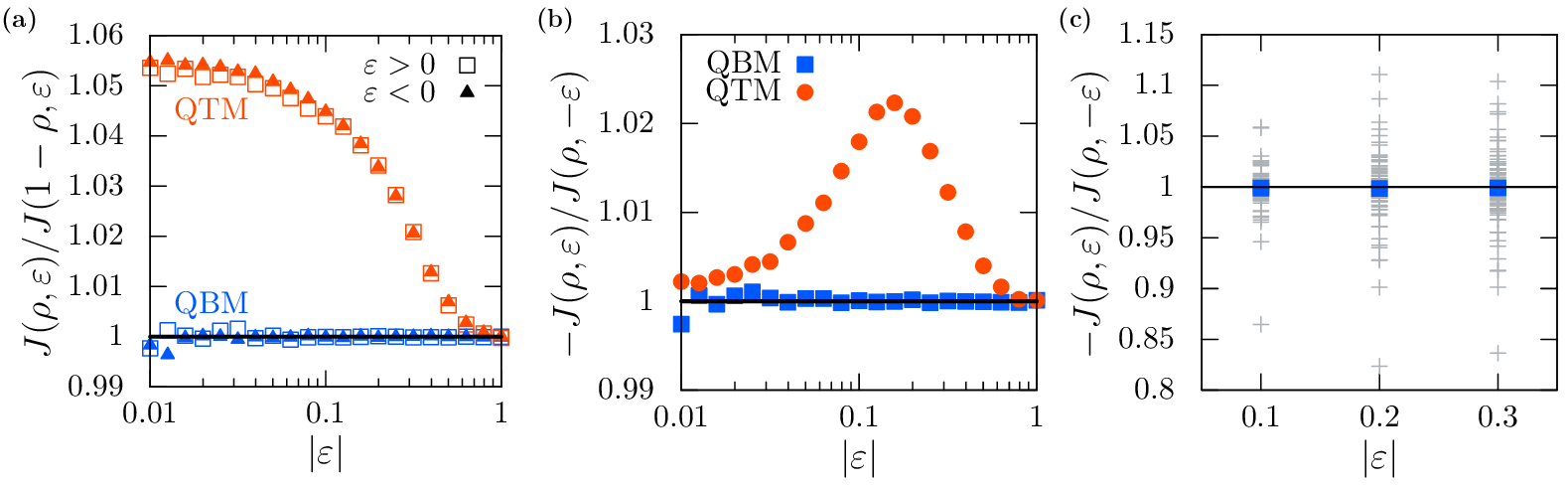}
    \caption{(a)--(c): Symmetry indicators vs the bias magnitude $|\varepsilon|$ ($L=100$, $T/T_g=2.5$, $\tau_c=1$).
    Panels (a) and (b) show results for a single quenched energy landscape, whereas panel (c) shows results for multiple independent landscapes.
    (a) Particle-hole symmetry ratio $J(\rho,\varepsilon)/J(1-\rho,\varepsilon)$ at $\rho=1/4$.
    Squares and triangles represent numerical results for $\varepsilon>0$ and for $\varepsilon<0$, respectively.
    Deviations from unity indicate particle-hole symmetry breaking.
    (b) Bias-reversal ratio $-J(\rho,\varepsilon)/J(\rho,-\varepsilon)$ at $\rho=1/2$.
    Symbols denote numerical results.
    Deviations from unity indicate bias-reversal asymmetry.
    (c) Bias-reversal ratio $-J(\rho,\varepsilon)/J(\rho,-\varepsilon)$ at $\rho=1/2$ in the QTM for various quenched energy landscapes.
    Cross symbols correspond to individual landscapes.
    Squares show the disorder-averaged result $-\Braket{J(\rho,\varepsilon)}_{\mathrm{dis}}/\Braket{J(\rho,-\varepsilon)}_{\mathrm{dis}}$, where $\Braket{\cdot}_{\mathrm{dis}}$ is the disorder average.}
    \label{fig: current-epsilon}
\end{figure*}

\subsection{Example 1: Quenched barrier model}
We consider the current in the system in which particles evolve according to the QBM [see Fig.~\ref{fig: model}(b)].
In this system, the bond-bias ratio $r_i(\varepsilon)$ is independent of $i$ because $E_{i,i+1}^{(\mathrm{b})}=E_{i+1,i}^{(\mathrm{b})}$,
and hence the particle-hole and bias-reversal symmetries hold.
Using a mean-field approximation and a perturbative expansion in $\varepsilon$,
the current can be obtained up to fourth order in $\varepsilon$ (see Appendix~\ref{appendix: D} for the derivation):
\begin{equation}
    \begin{split}
        J(\rho,\varepsilon)\cong&
        \varepsilon\frac{\rho(1-\rho)}{\mu}\\
        &-\varepsilon^3\frac{\left[\rho\left(1-\rho\right)\right]^2\sum_{i=1}^L\nu_i\nu_{i+1}}{L\mu}
        +O(\varepsilon^5),
    \end{split}
    \label{eq: current in QBM}
\end{equation}
where $\mu=\sum_{i=1}^Lw_{i,i+1}^{-1}/L$ is the sample average and $\nu_i=\sum_{j=0}^{L-1}(L-2j-1)w_{i+j,i+j+1}^{-1}/L\mu$.
The contributions of order $\varepsilon^2$ and $\varepsilon^4$ vanish.
The first-order term coincides with that of the homogeneous system,
whereas the third-order correction originates from the quenched disorder and is symmetric with respect to $\rho=1/2$.
Therefore, we obtain an approximate expression for the current up to fourth order in $\varepsilon$ and find results consistent with both the bias-reversal and particle-hole symmetries.

Figure~\ref{fig: current-density} compares the theoretical prediction [Eq.~\eqref{eq: current in QBM}] with numerical simulations.
For very weak bias, the current-density relation is well captured by the first-order term.
As the bias increases, the current decreases relative to the first-order prediction,
demonstrating the suppression of transport by heterogeneity.

We also numerically verify the particle-hole and bias-reversal symmetries in Fig.~\ref{fig: current-epsilon}.
For the QBM, the ratios $J(\rho,\varepsilon)/J(1-\rho,\varepsilon)$ and $-J(\rho,\varepsilon)/J(\rho,-\varepsilon)$ remain close to unity [Figs.~\ref{fig: current-epsilon}(a) and (b)].
Thus, both relations are indeed preserved in the QBM.
These results provide numerical confirmation of our conjecture.

\subsection{Example 2: Quenched trap model}
We next examine the particle-hole and bias-reversal symmetries in the QTM.
Figure~\ref{fig: current-epsilon}(a) shows that the ratio $J(\rho,\varepsilon)/J(1-\rho,\varepsilon)$ decreases monotonically with $|\varepsilon|$ and depends on the sign of $\varepsilon$.
For $|\varepsilon|\ll1$, the ratio deviates from unity, and the density dependence of the current exhibits the asymmetry around $\rho=1/2$.
As $|\varepsilon|$ increases, the density dependence of the current gradually approaches symmetry, but the approach itself differs between positive and negative biases because the current contains even-order contributions in $\varepsilon$ (see Appendix~\ref{appendix: E}).
For the extreme bias $|\varepsilon|=1$, it becomes symmetric about $\rho=1/2$.

Figure~\ref{fig: current-epsilon}(b) demonstrates that the bias-reversal symmetry is broken.  
In the linear-response regime, the bias-reversal symmetry holds trivially.
As $|\varepsilon|$ increases, nonlinear contributions induce an asymmetric response, such that $-J(\rho,\varepsilon)/J(\rho,-\varepsilon)\neq 1$.
With further increase in $|\varepsilon|$, this ratio gradually approaches unity, and for $|\varepsilon|=1$, the bias-reversal symmetry is restored.
Results for other densities are summarized in Appendix~\ref{appendix: F}, consistently showing an asymmetric response.

We now clarify the origin of the asymmetric response in the QTM.
For a single-particle system, the bias-reversal symmetry holds for all $\varepsilon$: $J(L^{-1},\varepsilon)=\varepsilon/\sum_{i=1}^L\tau_i$ \cite{PhysRevE.101.042133}, where $\tau_i=w_{i,i\pm1}^{-1}$ is the escape time at site $i$.
For an interacting particle system, as $|\varepsilon|$ increases, the bottleneck emerges and induces particle clogging.
A particle that has escaped the bottleneck may hop backward into it.
The asymmetry of escape rates across the bottleneck makes backward returns more likely for one bias direction, generating a sign-dependent clogging effect.
When the backward rate exceeds the forward rate, an escaped particle is more likely to return to the bottleneck, which impedes the current.
Thus, the sign-dependence of the return likelihood gives rise to an {\it interaction-induced asymmetric response}.
As the bias is further increased, the influence of backward hopping diminishes and the response approaches symmetry.

As shown in Fig.~\ref{fig: current-epsilon}(c), individual disorder realizations exhibit an asymmetric response, whereas the disorder-averaged current is nearly symmetric.
These realization-specific asymmetries cancel out upon averaging because the site disorder is drawn from a spatially unbiased distribution.
Each energy landscape contains incidental local structures that favor one bias direction or the other, but these directional preferences are equally likely to occur with opposite sign across different realizations.
Consequently, the positive and negative asymmetries compensate when averaged, yielding an almost symmetric disorder-averaged response.

\section{Conclusion}
We have established how environmental heterogeneity affects the bias-reversal symmetry of the current in the ASEP.
We showed that when the bond-bias ratio $p_i(\varepsilon)/q_{i+1}(\varepsilon)$ is uniform across all bonds, the bias-reversal symmetry holds, and in this case it is strictly equivalent to the particle-hole symmetry.
Conversely, when the ratio varies across bonds, this equivalence breaks down and at least one of the two symmetries is violated.
To substantiate these statements, we have considered the ASEP on a quenched random energy landscape as a concrete model.
When particles follow the QBM, we have obtained the approximate expression for the current up to fourth order in $\varepsilon$ and found results consistent with both the bias-reversal and particle-hole symmetries.
For a single particle system in the QTM, the bias-reversal symmetry holds,
whereas for the ASEP in the QTM, it is violated.
This violation arises from the interplay between the heterogeneous environment and many-body interactions.

Our framework provides a unified criterion that specifies when environmental heterogeneity preserves, and when it breaks, nonequilibrium transport symmetries in disordered exclusion processes.
Our theoretical predictions can be tested in colloidal systems where random energy landscapes are created by light fields~\cite{C2SM07102A}
and they provide insight into transport in biological ion channels and nanochannels used for drug delivery~\cite{Yang:2010aa,Zhao:2018aa}.
In particular, our findings shed light on weak rectification observed in biological ion channels, where transport proceeds through highly heterogeneous, narrow single-file channels.
The mechanism uncovered here---environmental heterogeneity combined with interaction-induced clogging---offers a mechanism for direction-dependent currents in single-file transport through narrow biological and artificial pores.

\begin{acknowledgments}
    This work was supported by JST SPRING, Grant No. JPMJSP2151.
    TA was supported by JSPS Grant-in-Aid for Scientific Research (Grant No. C JP21K033920).
\end{acknowledgments}

\appendix
\section{Condition for vanishing current in equilibrium}\label{appendix: A}
Here we discuss the condition under which no net current flows in the equilibrium system.
Let $\mathcal{P}_t(C)$ denote the probability that the system is in configuration $C$ at time $t$.
The time evolution of $\mathcal{P}_t(C)$ is governed by
\begin{equation}
    \frac{d\mathcal{P}_t(C)}{dt}
    =\sum_{C'}\left[W(C',C)\mathcal{P}_t(C')-W(C,C')\mathcal{P}_t(C)\right], \label{eqS:dp}
\end{equation}
where $W(C',C)$ is the transition rate from $C'$ to $C$.
In the long-time limit $t\rightarrow\infty$, the system reaches the steady state, so that $\mathcal{P}_t(C)\rightarrow \mathcal{P}_{\mathrm{eq}}(C)$.
Substituting $\mathcal{P}_{\mathrm{eq}}(C)$ into Eq.~\eqref{eqS:dp} gives
\begin{equation}
    \sum_{C'}\left[W(C',C)\mathcal{P_{\mathrm{eq}}}(C')-W(C,C')\mathcal{P}_{\mathrm{eq}}(C)\right]=0.
    \label{eqS: p_eq}
\end{equation}
Using the normalization condition $\sum_C\mathcal{P}_{\mathrm{eq}}(C)=1$, we obtain
\begin{equation}
    \mathcal{P}_{\mathrm{eq}}(C)=
    \frac{1}{Z}\prod_{i=1}^L\left[\sum_{j=0}^{L-1}\prod_{k=1}^jq_{i+k}(0)\prod_{l=j+1}^{L-1}p_{i+l}(0)\right]^{\eta_i(C)},
    \label{eqS: P_st}
\end{equation}
where $Z$ is the normalization constant and $\eta_i(C)$ is the occupation number at site $i$ in configuration $C$
($\eta_i=1$ if site $i$ is occupied and $\eta_i=0$ otherwise).
By substituting Eq.~\eqref{eqS: P_st} into the detailed balance relation $W(C',C)\mathcal{P}_{\mathrm{eq}}(C')=W(C,C')\mathcal{P}_{\mathrm{eq}}(C)$, we finally obtain the condition
\begin{equation}
    \prod_{i=1}^L\frac{w_{i,i-1}}{w_{i,i+1}}=1,
\end{equation}
which guarantees that no net current flows in equilibrium.

\section{Derivation of the linear-response relation}\label{appendix: B}
Here we derive the linear-response relation [Eq.~(7) in the main text] in the disordered ASEP.
The derivation is identical to that in Ref.~\cite{Hanney:2003aa}, and only the notation
and introduction of perturbations are adapted to our system.

\subsection{Second moment of the total integrated current in the equilibrium system}
We derive the second moment of the total integrated current in the equilibrium system.
We define the total integrated current $Q(t)$ as the net number of current-carrying events over all bonds up to time $t$: each $+1$ ($-1$) event changes $Q$ by $+1$ ($-1$).
Let $\mathcal{P}_t(C,Q)$ be the joint probability that the system is in configuration $C$ at time $t$ and that $Q(t)=Q$.
The time evolution of $\mathcal{P}_t(C,Q)$ is
\begin{equation}
    \begin{split}
        \frac{d\mathcal{P}_t(C,Q)}{dt}
        =&\sum_{C'}\left[W_{+1}(C',C)\mathcal{P}_t(C',Q-1)\right.\\
        &\left.+W_{-1}(C',C)P_t(C',Q+1)\right.\\
        &\left.-W(C,C')\mathcal{P}_t(C,Q)\right],
    \end{split}
\end{equation}
where $W_{\pm1}(C',C)$ is the transition rate from $C'$ to $C$ accompanied by a change $\pm1$ in $Q(t)$, and $W(C',C)=W_{+1}(C',C)+W_{-1}(C',C)$.
Using this equation, the time derivative of the second moment of the total integrated current is given by
\begin{equation}
    \begin{split}
        \frac{d\Braket{Q^2(t)}_0}{dt}
        =&\sum_{C,C'}\left[2\left(W_{+1}(C,C')-W_{-1}(C,C')\right)\mathcal{Q}_t(C)\right.\\
        &\left.+W(C,C')\mathcal{P}_t(C)\right],
    \end{split}
\end{equation}
where $\Braket{\cdot}_0$ denotes the thermal average, $\mathcal{P}_t(C)=\sum_{Q}\mathcal{P}_t(C,Q)$ and $\mathcal{Q}_t(C)=\sum_{Q}Q\mathcal{P}_t(C,Q)$.
$\mathcal{P}_t(C)$ is the probability of configuration $C$ at time $t$,
and $\mathcal{Q}_t(C)/\mathcal{P}_t(C)$ is the conditional average of the total current given configuration $C$ at time $t$.
The time evolutions of $\mathcal{P}_t(C)$ and $\mathcal{Q}_t(C)$ are Eq.~\eqref{eqS:dp} and
\begin{equation}
    \begin{split}
        \frac{d\mathcal{Q}_t(C)}{dt}
        &=\sum_{C'}\left[W(C',C)\mathcal{Q}_t(C')-W(C,C')\mathcal{Q}_t(C)\right.\\
        &\left.+\left(W_{+1}(C',C)-W_{-1}(C',C)\right)\mathcal{P}_t(C')\right].
        \label{eqS:dq}
    \end{split}
\end{equation}
Therefore, the steady probability $\mathcal{P}_t(C)\rightarrow\mathcal{P}_{\mathrm{eq}}(C)$ is given by Eq.~\eqref{eqS: P_st}.
Assuming $\mathcal{Q}_t(C)\to \mathcal{U}(C)\,t+\mathcal{R}(C)$ and separating coefficients of $t^1$ and $t^0$ in Eq.~\eqref{eqS:dq} give
\begin{equation}
    \sum_{C'} \left[ W(C',C) \mathcal{U}(C') - W(C,C') \mathcal{U}(C) \right] = 0, \label{eqS:uc_eq}
\end{equation}
\begin{equation}
    \begin{split}
        &\sum_{C'} \left[ W(C',C) \mathcal{R}(C') - W(C,C') \mathcal{R}(C) \right]\\
        =&\mathcal{U}(C) - \sum_{C'} \left[ W_{+1}(C',C) - W_{-1}(C',C) \right] \mathcal{P}_{\mathrm{eq}}(C').
    \end{split}
    \label{eqS:rc_eq}
\end{equation}
Equation~\eqref{eqS:uc_eq} is the same as the stationary equation of $\mathcal{P}_{\mathrm{eq}}(C)$ [Eq.~\eqref{eqS: p_eq}].
Thus, $\mathcal{U}(C)=A\mathcal{P}_{\mathrm{eq}}(C)$ for some constant $A$.
Summing both sides of Eq.~\eqref{eqS:rc_eq} concerning $C$, we obtain $A=\sum_C\mathcal{U}(C)=0$, i.e., $\mathcal{U}(C)=0$.
Therefore, Eq.~\eqref{eqS:uc_eq} becomes
\begin{equation}
    \begin{split}
        &\sum_{C'} \left[ W(C',C) \mathcal{R}(C') - W(C,C') \mathcal{R}(C) \right]\\
        =&- \sum_{C'} \left[ W_{+1}(C',C) - W_{-1}(C',C) \right] \mathcal{P}_{\mathrm{eq}}(C').
    \end{split}
    \label{eqS:rc_eq2}
\end{equation}

Finally, the second moment of the total integrated current in equilibrium is obtained as
\begin{equation}
    \begin{split}
        \lim_{t\rightarrow\infty}\frac{d\Braket{Q^2(t)}_{0}}{dt}
        =&\sum_{C,C'}\left[2\left(W_{+1}(C,C')-W_{-1}(C,C')\right)\mathcal{R}(C)\right.\\
        &\left.+W(C,C')\mathcal{P}_{\mathrm{eq}}(C)\right].
    \end{split}
\end{equation}

\begin{widetext}

\subsection{Average of the total integrated current in the nonequilibrium system}
We derive the average of the total integrated current in the nonequilibrium system.
Let $P_t(C)$ be the probability that the system is in configuration $C$ at time $t$.
The time evolution of $P_t(C)$ is
\begin{equation}
    \begin{split}
        \frac{dP_t(C)}{dt}=
        &\sum_{C'}\left[\left(\left(1+\varepsilon\right)W_{+1}(C',C)+\left(1-\varepsilon\right)W_{-1}(C',C)\right)P_t(C')\right.\\
        &\left.-\left(\left(1+\varepsilon\right)W_{+1}(C,C')+\left(1-\varepsilon\right)W_{-1}(C,C')\right)P_t(C)\right],
    \end{split}
    \label{eqS: dP noneq}
\end{equation}
where $\varepsilon$ is a bias parameter ($-1\leq\varepsilon\leq1$) and $W_{\pm1}(C',C)$ is the transition rate from configuration $C'$ and $C$ in the equilibrium system ($\varepsilon=0$).
In the long-time limit, the system reaches the steady state and $P_t(C) \to P(C)$.
We perform the perturbative expansion of $P(C)$ in powers of $\varepsilon$,
\begin{equation}
    P(C)=\sum_{n=0}^\infty\varepsilon^nP^{(n)}(C).
\end{equation}
Substituting this expansion into Eq.~\eqref{eqS: dP noneq}, the equations of $P^{(0)}(C)$ and $P^{(1)}(C)$ are
\begin{equation}
    \sum_{C'}\left[W(C',C)P^{(0)}(C')-W(C,C')P^{(0)}(C)\right]=0,
    \label{eqS: P0 noneq}
\end{equation}
\begin{equation}
    \begin{split}
        &\sum_{C'}\left[W(C',C)P^{(1)}(C')-W(C,C')P^{(1)}(C)\right]\\
        =&-\sum_{C'}\left[W_{+1}(C',C)-W_{-1}(C',C)\right]P^{(0)}(C')
        +\sum_{C'}\left[W_{+1}(C,C')-W_{-1}(C,C')\right]P^{(0)}(C),
    \end{split}
    \label{eqS: P1 noneq}
\end{equation}
respectively.
Because $P^{(0)}(C)$ satisfies the normalization $\sum_{C}P^{(0)}(C)=1$ and Eq.~\eqref{eqS: P0 noneq} is the same as Eq.~\eqref{eqS: p_eq}, we have $P^{(0)}(C)=\mathcal{P}_{\mathrm{eq}}(C)$.
Using detailed balance relation $W(C',C)P^{(0)}(C')=W(C,C')P^{(0)}(C)$, Eq.~\eqref{eqS: P1 noneq} is identical to Eq.~\eqref{eqS:rc_eq2}, that is, $P^{(1)}(C)=2\mathcal{R}(C)$.

The average of the integrated total current satisfies
\begin{equation}
    \lim_{t\rightarrow\infty}\frac{d\Braket{Q(t)}_{\varepsilon}}{dt}
        =\sum_{C,C'}\left[\left(1+\varepsilon\right)W_{+1}(C,C')-\left(1-\varepsilon\right)W_{-1}(C,C')\right]P(C),
\end{equation}
where $\Braket{\cdot}_{\varepsilon}$ is the ensemble average under $\varepsilon$.
The first-order term of $\Braket{Q(t)}_{\varepsilon}$ is represented
\begin{equation}
    \lim_{t\rightarrow\infty}\frac{d\Braket{Q(t)}_\varepsilon}{dt}\sim\varepsilon\sum_{C,C'}\left[W(C,C')P^{(0)}(C)+\left(W_{+1}(C,C')-W_{-1}(C,C')\right)P^{(1)}(C)\right]\quad(\varepsilon\rightarrow0).
\end{equation}
Using the relations $P^{(0)}=\mathcal{P}_{\mathrm{eq}}(C)$ and $P^{(1)}=2\mathcal{R}(C)$, we obtain the linear-response relation
\begin{equation}
    \lim_{t\rightarrow\infty}\frac{d\Braket{Q(t)}_\varepsilon}{dt}\sim\varepsilon\lim_{t\rightarrow\infty}\frac{d\Braket{Q^2(t)}_{0}}{dt}.
\end{equation}
In the steady state, the particle current $J(\rho,\varepsilon)$---the average number of particles crossing a bond per unit time at density $\rho$ under the bias parameter $\varepsilon$---does not depend on the bond, and hence $\lim_{t\rightarrow\infty}d\Braket{Q(t)}_{\varepsilon}/dt=LJ(\rho,\varepsilon)$.
Therefore,
\begin{equation}
    J(\rho,\varepsilon)
    \sim \frac{\varepsilon}{L}\lim_{t\to\infty}\frac{d\Braket{Q(t)^2}_0}{dt}\quad(\varepsilon\rightarrow0).
    \label{eqS: linear response relation}
\end{equation}

\section{Inversion-correspondence condition for the next-order contribution to be of order $\varepsilon^3$}\label{appendix: C}
We prove that the next-order contribution of the current is at order $\varepsilon^3$ under the inversion-correspondence condition.
In this condition, there exists $k\in\{1,\dots,L\}$ such that for $i\in\{0,\dots,L-1\}$, either $p_{k+i}(\varepsilon)=q_{k-i}(-\varepsilon)$ or $p_{k+i}(\varepsilon)=q_{k-i+1}(-\varepsilon)$ holds.

\subsection{Case of $p_{k+i}(\varepsilon)=q_{k-i}(-\varepsilon)$}
The current under $\varepsilon$ at the bond between site $k+i$ and $k+i+1$ is given by
\begin{equation}
    J(\rho,\varepsilon)
    =p_{k+i}(\varepsilon)\Braket{\eta_{k+i}(1-\eta_{k+i+1})}_{\varepsilon}-q_{k+i+1}(\varepsilon)\Braket{\eta_{k+i+1}(1-\eta_{k+i})}_{\varepsilon}.
    \label{eqS: current k+i k+i+1}
\end{equation}
In contrast, the current under the reversed bias $-\varepsilon$ at the bond between site $k-i-1$ and $k-i$ is represented by
\begin{equation}
    \begin{split}
        J(\rho,-\varepsilon)
        =&p_{k-i-1}(-\varepsilon)\Braket{\eta_{k-i-1}(1-\eta_{k-i})}_{-\varepsilon}-q_{k-i}(-\varepsilon)\Braket{\eta_{k-i}(1-\eta_{k-i-1})}_{-\varepsilon}\\
        =&q_{k+i+1}(\varepsilon)\Braket{\eta_{k-i-1}(1-\eta_{k-i})}_{-\varepsilon}-p_{k+i}(\varepsilon)\Braket{\eta_{k-i}(1-\eta_{k-i-1})}_{-\varepsilon},
    \end{split}
\end{equation}
where in the second equality we used $p_{k+i}(\varepsilon)=q_{k-i}(-\varepsilon)$.

The stationarity equations for one- and two-point functions under $+\varepsilon$ read
\begin{equation}
    \begin{split}
        &p_{k+i}(\varepsilon)\Braket{\eta_{k+i}(1-\eta_{k+i+1})}_{\varepsilon}-q_{k+i+1}(\varepsilon)\Braket{\eta_{k+i+1}(1-\eta_{k+i})}_{\varepsilon}\\
        =&p_{k+i+1}(\varepsilon)\Braket{\eta_{k+i+1}(1-\eta_{k+i+2})}_{\varepsilon}-q_{k+i+2}(\varepsilon)\Braket{\eta_{k+i+2}(1-\eta_{k+i+1})}_{\varepsilon},
    \end{split}
    \label{eqS: rho equation for +}
\end{equation}
while those under $-\varepsilon$ are
\begin{equation}
    \begin{split}
        &q_{k+i+1}(\varepsilon)\Braket{\eta_{k-i-1}(1-\eta_{k-i})}_{-\varepsilon}-p_{k+i}(\varepsilon)\Braket{\eta_{k-i}(1-\eta_{k-i-1})}_{-\varepsilon}\\
        =&q_{k+i+2}(\varepsilon)\Braket{\eta_{k-i-2}(1-\eta_{k-i-1})}_{-\varepsilon}-p_{k+i+1}(\varepsilon)\Braket{\eta_{k-i-1}(1-\eta_{k-i-2})}_{-\varepsilon}.
    \end{split}
    \label{eqS: rho equation for -}
\end{equation}
Equations~\eqref{eqS: rho equation for +} and \eqref{eqS: rho equation for -} are identical, and therefore
\begin{equation}
    \Braket{\eta_{k+i}}_{\varepsilon}=\Braket{\eta_{k-i}}_{-\varepsilon},\quad
    \Braket{\eta_{k+i}\eta_{k+i+1}}_{\varepsilon}=\Braket{\eta_{k-i}\eta_{k-i-1}}_{-\varepsilon}.
\end{equation}
Using these relations in the expression for $J(\rho,-\varepsilon)$ yields
\begin{equation}
    J(\rho,-\varepsilon)=q_{k+i+1}(\varepsilon)\Braket{\eta_{k+i+1}(1-\eta_{k+i})}_{\varepsilon}-p_{k+i}(\varepsilon)\Braket{\eta_{k+i}(1-\eta_{k+i+1})}_\varepsilon
    =-J(\rho,\varepsilon),
\end{equation}
which proves the symmetric response for the case $p_{k+i}(\varepsilon)=q_{k-i}(-\varepsilon)$.

\subsection{Case of $p_{k+i}(\varepsilon)=q_{k-i+1}(-\varepsilon)$}
Similarly to the case $p_{k+i}(\varepsilon)=q_{k-i}(-\varepsilon)$, we derive the symmetric response $J(\rho,\varepsilon)=-J(\rho,-\varepsilon)$ for $p_{k+i}(\varepsilon)=q_{k-i+1}(-\varepsilon)$.
The current under $\varepsilon$ at the bond between site $k+i$ and $k+i+1$ is given by Eq.~\eqref{eqS: current k+i k+i+1}, whereas under the reversed bias $-\varepsilon$ at the bond between sites $k-i$ and $k-i+1$ it is
\begin{equation}
    \begin{split}
        J(\rho,-\varepsilon)=&
        p_{k-i}(-\varepsilon)\Braket{\eta_{k-i}(1-\eta_{k-i+1})}_{-\varepsilon}-q_{k-i+1}(-\varepsilon)\Braket{\eta_{k-i+1}(1-\eta_{k-i})}_{-\varepsilon}\\
        =&q_{k+i+1}(\varepsilon)\Braket{\eta_{k-i}(1-\eta_{k-i+1})}_{-\varepsilon}-p_{k+i}(\varepsilon)\Braket{\eta_{k-i+1}(1-\eta_{k-i})}_{-\varepsilon},
    \end{split}
\end{equation}
where in the second equality we used $p_{k+i}(\varepsilon)=q_{k-i+1}(-\varepsilon)$.

The stationarity equations under $+\varepsilon$ are the same as Eq.~\eqref{eqS: rho equation for +}, while those under $-\varepsilon$ read
\begin{equation}
    \begin{split}
        &q_{k+i+1}(\varepsilon)\Braket{\eta_{k-i}(1-\eta_{k-i+1})}_{-\varepsilon}-p_{k+i}(\varepsilon)\Braket{\eta_{k-i+1}(1-\eta_{k-i})}_{-\varepsilon}\\
        =&q_{k+i+2}(\varepsilon)\Braket{\eta_{k-i-1}(1-\eta_{k-i})}_{-\varepsilon}-p_{k+i+1}(\varepsilon)\Braket{\eta_{k-i}(1-\eta_{k-i-1})}_{-\varepsilon}.
    \end{split}
    \label{eqS: rho equation for -2}
\end{equation}
Equations~\eqref{eqS: rho equation for +} and \eqref{eqS: rho equation for -2} are identical under the identifications
\begin{equation}
    \Braket{\eta_{k+i}}_{\varepsilon}=\Braket{\eta_{k-i+1}}_{-\varepsilon},\qquad
    \Braket{\eta_{k+i}\eta_{k+i+1}}_{\varepsilon}=\Braket{\eta_{k-i+1}\eta_{k-i}}_{-\varepsilon}.
    \label{eqS: map shift}
\end{equation}
Using these relations, we obtain
\begin{equation}
    \begin{split}
        J(\rho,-\varepsilon)
        =&\,q_{k+i+1}(\varepsilon)\Braket{\eta_{k+i+1}(1-\eta_{k+i})}_{\varepsilon}
        -p_{k+i}(\varepsilon)\Braket{\eta_{k+i}(1-\eta_{k+i+1})}_{\varepsilon}\\
        =&\,-J(\rho,\varepsilon),
    \end{split}
\end{equation}
which proves the symmetric response also for the case $p_{k+i}(\varepsilon)=q_{k-i+1}(-\varepsilon)$.

\end{widetext}

\section{Derivation of the current in the quenched barrier model}\label{appendix: D}
We consider the disordered ASEP in which particles perform the quenched barrier model (QBM).
Using the mean-field approximation and the perturbative expansion, we derive the current up to fourth order in $\varepsilon$.

The current between site $i$ and $i+1$ is given by
\begin{equation}
    \begin{split}
        J_{i,i+1}(\rho,\varepsilon)=&\frac{1+\varepsilon}{2}w_{i,i+1}\Braket{\eta_i(1-\eta_{i+1})}_{\varepsilon}\\
        &-\frac{1-\varepsilon}{2}w_{i+1,i}\Braket{\eta_{i+1}(1-\eta_i)}_{\varepsilon},
    \end{split}
\end{equation}
where $\Braket{\cdot}_{\varepsilon}$ denotes the steady-state average at bias $\varepsilon$, and $\eta_i$ is the occupation number at site $i$.
The continuity equation $\partial_t\Braket{\eta_i(t)}_{\varepsilon}=J_{i-1,i}(\rho,\varepsilon)-J_{i,i+1}(\rho,\varepsilon)$ implies that in the steady state ($\partial_t\Braket{\eta_i(t)}_{\varepsilon}=0$) the current is independent of $i$:
$J(\rho,\varepsilon)=J_{i-1,i}(\rho,\varepsilon)=J_{i,i+1}(\rho,\varepsilon)$.
For the QBM, we have $w_{i,i+1}=w_{i+1,i}$ for all $i$, hence
\begin{equation}
    J(\rho,\varepsilon)=w_{i,i+1}\left[\frac{1+\varepsilon}{2}\Braket{\eta_i}_{\varepsilon}-\frac{1-\varepsilon}{2}\Braket{\eta_{i+1}}_{\varepsilon}-\varepsilon\Braket{\eta_i\eta_{i+1}}_{\varepsilon}\right].
    \label{eqS: current}
\end{equation}
Dividing Eq.~\eqref{eqS: current} by $w_{i,i+1}$ and summing over $i$ (with periodic boundaries) yields
\begin{equation}
    \begin{split}
        J(\rho,\varepsilon)
        =&\frac{\varepsilon}{\sum_{i=1}^Lw_{i,i+1}^{-1}}\left[\sum_{i=1}^L\Braket{\eta_i}_{\varepsilon}-\sum_{i=1}^L\Braket{\eta_i\eta_{i+1}}_{\varepsilon}\right]\\
        =&\frac{\varepsilon}{L\mu}\left[N-\sum_{i=1}^L\Braket{\eta_i\eta_{i+1}}_{\varepsilon}\right],
    \end{split}
\end{equation}
where $\mu=\frac{1}{L}\sum_{i=1}^L w_{i,i+1}^{-1}$ is the sample average of the inverse rates, and we used particle-number conservation $\sum_{i=1}^L\Braket{\eta_i}_{\varepsilon}=N$.

We expand $\Braket{\eta_i}_{\varepsilon}$ and $\Braket{\eta_i\eta_{i+1}}_{\varepsilon}$ in powers of $\varepsilon$:
\begin{equation}
    \Braket{\eta_i}_{\varepsilon}=\sum_{n=0}^{\infty}\varepsilon^n\rho_i^{(n)},\quad
    \Braket{\eta_i\eta_{i+1}}_{\varepsilon}=\sum_{n=0}^{\infty}\varepsilon^n\sigma_{i,i+1}^{(n)},
\end{equation}
where $\rho_i^{(n)}$ and $\sigma_{i,i+1}^{(n)}$ are the $n$th-order contributions.
In equilibrium ($\varepsilon=0$), all configurations with $N$ particles are equiprobable, $P(C)=\binom{L}{N}^{-1}$, so
\begin{equation}
    \rho_i^{(0)}=\frac{N}{L},\quad
    \sigma_{i,i+1}^{(0)}=\frac{N(N-1)}{L(L-1)}.
\end{equation}
Since the first-order term of the current depends only on $\sigma_{i,i+1}^{(0)}$, we obtain
\begin{equation}
    J(\rho,\varepsilon)
    =\frac{\varepsilon}{\mu}\rho(1-\rho)
    -\frac{1}{L\mu}\sum_{n=2}^{\infty}\varepsilon^n\sum_{i=1}^L\sigma_{i,i+1}^{(n-1)}
    \label{eqS: current for QBM}
\end{equation}
in the thermodynamic limit $N/L\to\rho$ ($L\to\infty$ and $N\to\infty$).
Therefore, the density dependence of the linear response of the current coincides with that of the homogeneous system.

\subsection{First-order corrections}
For the QBM, we can solve Eq.~\eqref{eqS: P1 noneq} explicitly:
\begin{equation}
    P^{(1)}(i_1,i_2,\dots,i_N)
    =\frac{\sum_{n=1}^N\sum_{j=0}^{L-1}(L-1-2j)w_{i_n+j,i_n+j+1}^{-1}}{\binom{L}{N}\sum_{j=1}^Lw_{j,j+1}^{-1}},
\end{equation}
where $i_n$ denotes the position of particle $n$.
This gives
\begin{align}
    &\rho_i^{(1)}=\frac{N(L-N)}{L(L-1)}\frac{\sum_{j=0}^{L-1}(L-2j-1)w_{i+j,i+j+1}^{-1}}{\sum_{j=1}^Lw_{j,j+1}^{-1}},\label{eqS: rho1}\\
    &\sigma_{i,i+1}^{(1)}=\frac{2N(N-1)(L-N)}{L(L-1)(L-2)}\frac{\sum_{j=1}^{L-1}(L-2j)w_{i+j,i+j+1}^{-1}}{\sum_{j=1}^Lw_{j,j+1}^{-1}}.
\end{align}
Because $\sum_{i=1}^L \sigma_{i,i+1}^{(1)}=0$, the second-order response of the current vanishes.

\subsection{Second-order corrections}
Solving the full equation for $P^{(2)}(C)$ is difficult analytically.
We therefore employ a mean-field approximation (MFA):
\begin{equation}
    \sigma_{i,i+1}^{(2)}\cong\rho_i^{(2)}\rho_{i+1}^{(0)}+\rho_i^{(0)}\rho_{i+1}^{(2)}+\rho_i^{(1)}\rho_{i+1}^{(1)}.
    \label{eqS: sigma2 MFA}
\end{equation}
Hence it suffices to determine $\rho_i^{(2)}$, which satisfies
\begin{align}
    &\rho_i^{(2)}-\rho_{i+1}^{(2)}=2\sigma_{i,i+1}^{(1)}-\rho_i^{(1)}-\rho_{i+1}^{(1)},\\
    &\sum_{i=1}^L\rho_i^{(2)}=0.
\end{align}
The solution is
\begin{equation}
    \begin{split}
        \rho_i^{(2)}
        =&\frac{\sum_{j=1}^{L-1}(L-1)\left[2\sigma_{i+j-1,i+j}^{(1)}-\rho_{i+j-1}^{(1)}-\rho_{i+j}^{(1)}\right]}{L}\\
        =&\frac{2N-L}{L(L-2)}\left[L\rho_i^{(1)}-2\sum_{j=1}^Lj\rho_{i+j}^{(1)}\right],
    \end{split}
\end{equation}
where in the second equality we used $\sigma_{i,i+1}^{(1)}=\frac{N-1}{L-2}\big(\rho_i^{(1)}+\rho_{i+1}^{(1)}\big)$.

\subsection{Third- and fourth-order responses of the current}
From Eq.~\eqref{eqS: current for QBM}, the third- and fourth-order responses are proportional to $\sum_{i=1}^L \sigma_{i,i+1}^{(2)}$ and $\sum_{i=1}^L \sigma_{i,i+1}^{(3)}$, respectively.
Using Eqs.~\eqref{eqS: rho1} and \eqref{eqS: sigma2 MFA},
\begin{equation}
    \sum_{i=1}^L\sigma_{i,i+1}^{(2)}\cong\sum_{i=1}^L\rho_i^{(1)}\rho_{i+1}^{(1)}=\left[\rho(1-\rho)\right]^2\sum_{i=1}^L\nu_i\nu_{i+1},
\end{equation}
with
\begin{equation}
    \nu_i
    =\frac{\sum_{j=0}^{L-1}(L-2j-1)w_{i+j,i+j+1}^{-1}}{L\mu}.
\end{equation}
For the third-order two-site term we use the MFA
$\sigma_{i,i+1}^{(3)}\cong\sum_{n=0}^3 \rho_i^{(n)}\rho_{i+1}^{(3-n)}$, which gives
\begin{equation}
    \begin{split}
        \sum_{i=1}^L\sigma_{i,i+1}^{(3)}
        \cong&\sum_{i=1}^L\left[\rho_i^{(1)}\rho_{i+1}^{(2)}+\rho_i^{(2)}\rho_{i+1}^{(1)}\right]\\
        =&-\frac{2(2N-L)}{L(L-2)}\sum_{i=1}^L\rho_i^{(1)}\left(L\rho_i^{(1)}+2\sum_{j=1}^{L-1}j\rho_{i+j}^{(1)}\right).
    \end{split}
\end{equation}
Using $\sum_{j=1}^{L-1}\rho_{i+j}^{(1)}=-\rho_i^{(1)}$ (periodicity and zero sum), we obtain
\begin{equation}
    \sum_{i=1}^L\sigma_{i,i+1}^{(3)}
    \cong\frac{2(2N-L)}{L(L-2)}\sum_{i=1}^L\rho_i^{(1)}\sum_{j=1}^{L-1}(L-2j)\rho_{i+j}^{(1)}.
\end{equation}
Comparing the pair $(i=m,\,j=n-m>0)$ with $(i=n,\,j=m-n+L)$, we have
\begin{equation}
    (L-2(n-m))\rho_{m}^{(1)}\rho_{n}^{(1)}+(L-2(m-n+L))\rho_{n}^{(1)}\rho_{m+L}^{(1)}=0,
\end{equation}
and since $\rho_{m+L}^{(1)}=\rho_m^{(1)}$, it follows that $\sum_{i=1}^L \sigma_{i,i+1}^{(3)}=0$.
Therefore, the current up to fourth order in $\varepsilon$ is
\begin{equation}
    J(\rho,\varepsilon)
    \cong\frac{\varepsilon}{\mu}\rho(1-\rho)
    -\varepsilon^3\frac{\left[\rho(1-\rho)\right]^2\sum_{i=1}^L\nu_i\nu_{i+1}}{L\mu}
    +O(\varepsilon^5).
\end{equation}

\section{Dependence of density asymmetry on sign of applied bias}\label{appendix: E}
We show that the ratio $J(\rho,\varepsilon)/J(1-\rho,\varepsilon)$ depends on the sign of the applied bias when the asymmetric response $J(\rho,\varepsilon)\neq -J(\rho,-\varepsilon)$ occurs.
We expand the current as
\begin{equation}
    J(\rho,\varepsilon)=A(\rho)\varepsilon+B(\rho)\varepsilon^2,
\end{equation}
where $A(\rho)$ and $B(\rho)$ represent the density-dependent coefficients of the linear and quadratic contributions, respectively.

The ratio $J(\rho,\varepsilon)/J(1-\rho,\varepsilon)$ becomes
\begin{equation}
    \begin{split}
        \frac{J(\rho,\varepsilon)}{J(1-\rho,\varepsilon)}
        \sim&\frac{A(\rho)+B(\rho)\varepsilon}{A(1-\rho)}\left(1+\frac{B(1-\rho)}{A(1-\rho)}\varepsilon\right)^{-1}\\
        \sim&\frac{A(\rho)}{A(1-\rho)}\\
        &+\frac{B(\rho)A(1-\rho)-A(\rho)B(1-\rho)}{A(1-\rho)^2}\varepsilon+O(\varepsilon^2).
    \end{split}
\end{equation}
Thus, the ratio is independent of the sign of $\varepsilon$ if and only if all odd-order terms vanish.
When the odd-order terms become zero, the ratio does not depend on the sign of the applied bias.
At the lowest order, the coefficient of $\varepsilon$ becomes zero under either of the following conditions: (i) $B(\rho)=0$, that is, the current is an odd function of $\varepsilon$, implying a symmetric response; (ii)
\begin{equation}
    \frac{A(\rho)}{A(1-\rho)}=\frac{B(\rho)}{B(1-\rho)},
\end{equation}
that is, the density dependences of $A(\rho)$ and $B(\rho)$ are identical.
For the QTM, neither condition is satisfied:
(i) $B(\rho)$ is nonzero; and
(ii) the density dependence of $B(\rho)$ differs from that of $A(\rho)$.
Consequently, the linear term of the ratio is nonvanishing, and the ratio $J(\rho,\varepsilon)/J(1-\rho,\varepsilon)$ necessarily depends on the sign of the applied bias, consistent with the numerical results in the main text.

\section{Additional numerical results}\label{appendix: F}
\begin{figure}[h]
    \centering
    \includegraphics[width=8cm]{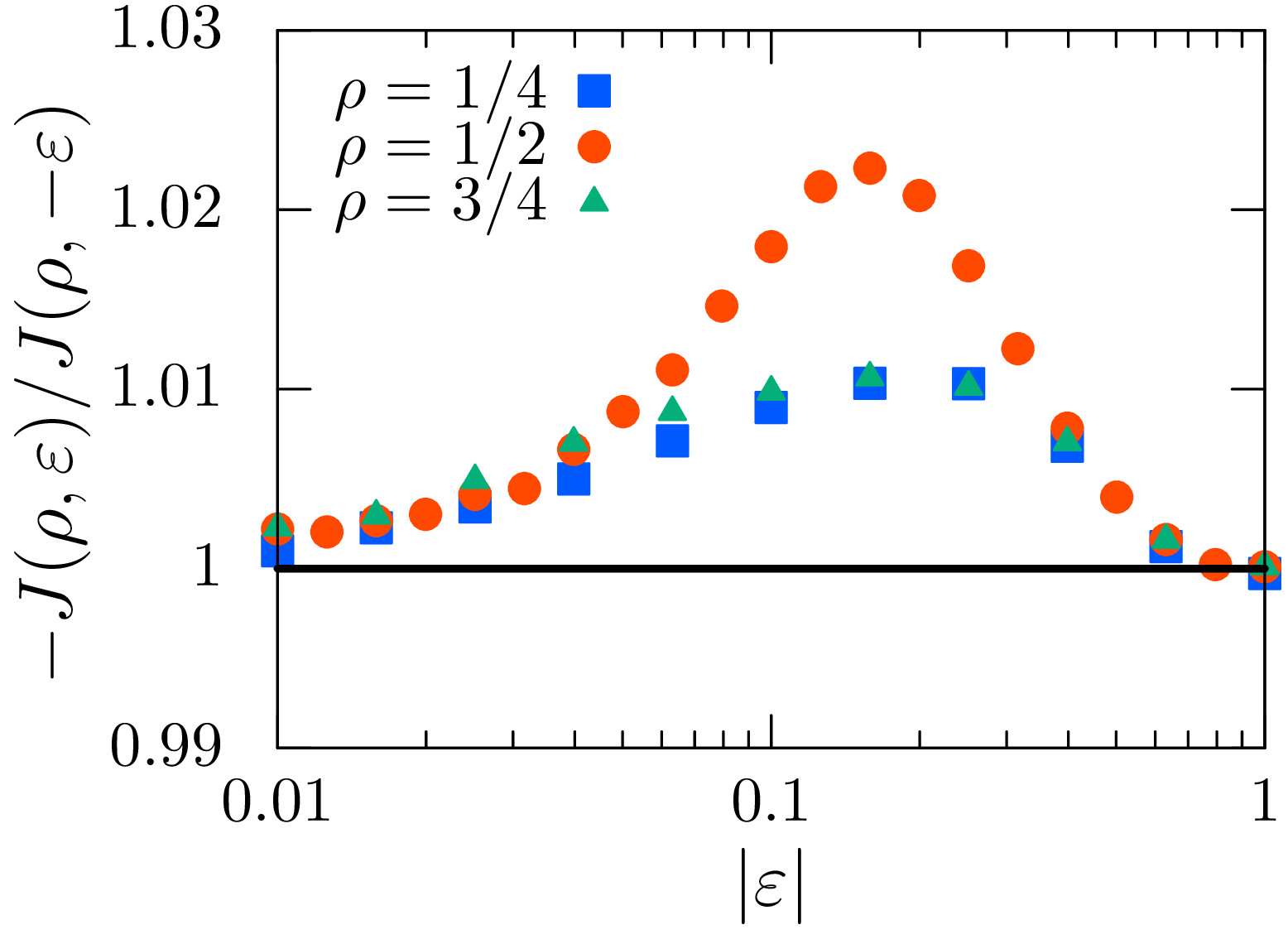}
    \caption{Bias-reversal ratio $-J(\rho,\varepsilon)/J(\rho,-\varepsilon)$ vs the bias magnitude $|\varepsilon|$ in the QTM ($L=100,~T/T_g=2.5,~\tau_c=1$).}
    \label{fig: Bias reversal}
\end{figure}

Figure~\ref{fig: Bias reversal} shows that the bias-reversal symmetry is broken at densities other than $\rho=1/2$.
The ratios $-J(\rho,\varepsilon)/J(\rho,-\varepsilon)$ for $\rho=1/4$ and $3/4$ are closer to $1$ than the ratio for $\rho=1/2$.

\bibliography{ASEP}

\begin{thebibliography}{47}%
\makeatletter
\providecommand \@ifxundefined [1]{%
 \@ifx{#1\undefined}
}%
\providecommand \@ifnum [1]{%
 \ifnum #1\expandafter \@firstoftwo
 \else \expandafter \@secondoftwo
 \fi
}%
\providecommand \@ifx [1]{%
 \ifx #1\expandafter \@firstoftwo
 \else \expandafter \@secondoftwo
 \fi
}%
\providecommand \natexlab [1]{#1}%
\providecommand \enquote  [1]{``#1''}%
\providecommand \bibnamefont  [1]{#1}%
\providecommand \bibfnamefont [1]{#1}%
\providecommand \citenamefont [1]{#1}%
\providecommand \href@noop [0]{\@secondoftwo}%
\providecommand \href [0]{\begingroup \@sanitize@url \@href}%
\providecommand \@href[1]{\@@startlink{#1}\@@href}%
\providecommand \@@href[1]{\endgroup#1\@@endlink}%
\providecommand \@sanitize@url [0]{\catcode `\\12\catcode `\$12\catcode `\&12\catcode `\#12\catcode `\^12\catcode `\_12\catcode `\%12\relax}%
\providecommand \@@startlink[1]{}%
\providecommand \@@endlink[0]{}%
\providecommand \url  [0]{\begingroup\@sanitize@url \@url }%
\providecommand \@url [1]{\endgroup\@href {#1}{\urlprefix }}%
\providecommand \urlprefix  [0]{URL }%
\providecommand \Eprint [0]{\href }%
\providecommand \doibase [0]{https://doi.org/}%
\providecommand \selectlanguage [0]{\@gobble}%
\providecommand \bibinfo  [0]{\@secondoftwo}%
\providecommand \bibfield  [0]{\@secondoftwo}%
\providecommand \translation [1]{[#1]}%
\providecommand \BibitemOpen [0]{}%
\providecommand \bibitemStop [0]{}%
\providecommand \bibitemNoStop [0]{.\EOS\space}%
\providecommand \EOS [0]{\spacefactor3000\relax}%
\providecommand \BibitemShut  [1]{\csname bibitem#1\endcsname}%
\let\auto@bib@innerbib\@empty
%</preamble>
\bibitem [{\citenamefont {Spitzer}(1970)}]{SPITZER1970246}%
  \BibitemOpen
  \bibfield  {author} {\bibinfo {author} {\bibfnamefont {F.}~\bibnamefont {Spitzer}},\ }\bibfield  {title} {\bibinfo {title} {Interaction of markov processes},\ }\href {https://doi.org/https://doi.org/10.1016/0001-8708(70)90034-4} {\bibfield  {journal} {\bibinfo  {journal} {Advances in Mathematics}\ }\textbf {\bibinfo {volume} {5}},\ \bibinfo {pages} {246} (\bibinfo {year} {1970})}\BibitemShut {NoStop}%
\bibitem [{\citenamefont {Derrida}(1998)}]{DERRIDA199865}%
  \BibitemOpen
  \bibfield  {author} {\bibinfo {author} {\bibfnamefont {B.}~\bibnamefont {Derrida}},\ }\bibfield  {title} {\bibinfo {title} {An exactly soluble non-equilibrium system: The asymmetric simple exclusion process},\ }\href {https://doi.org/https://doi.org/10.1016/S0370-1573(98)00006-4} {\bibfield  {journal} {\bibinfo  {journal} {Physics Reports}\ }\textbf {\bibinfo {volume} {301}},\ \bibinfo {pages} {65} (\bibinfo {year} {1998})}\BibitemShut {NoStop}%
\bibitem [{\citenamefont {Derrida}(2007)}]{Derrida_2007}%
  \BibitemOpen
  \bibfield  {author} {\bibinfo {author} {\bibfnamefont {B.}~\bibnamefont {Derrida}},\ }\bibfield  {title} {\bibinfo {title} {Non-equilibrium steady states: fluctuations and large deviations of the density and of the current},\ }\href {https://doi.org/10.1088/1742-5468/2007/07/P07023} {\bibfield  {journal} {\bibinfo  {journal} {Journal of Statistical Mechanics: Theory and Experiment}\ }\textbf {\bibinfo {volume} {2007}},\ \bibinfo {pages} {P07023} (\bibinfo {year} {2007})}\BibitemShut {NoStop}%
\bibitem [{\citenamefont {De~Masi}\ and\ \citenamefont {Ferrari}(1985)}]{De-Masi:1985aa}%
  \BibitemOpen
  \bibfield  {author} {\bibinfo {author} {\bibfnamefont {A.}~\bibnamefont {De~Masi}}\ and\ \bibinfo {author} {\bibfnamefont {P.~A.}\ \bibnamefont {Ferrari}},\ }\bibfield  {title} {\bibinfo {title} {Self-diffusion in one-dimensional lattice gases in the presence of an external field},\ }\href {https://doi.org/10.1007/BF01010480} {\bibfield  {journal} {\bibinfo  {journal} {Journal of Statistical Physics}\ }\textbf {\bibinfo {volume} {38}},\ \bibinfo {pages} {603} (\bibinfo {year} {1985})}\BibitemShut {NoStop}%
\bibitem [{\citenamefont {Derrida}\ and\ \citenamefont {Mallick}(1997)}]{Derrida:1997aa}%
  \BibitemOpen
  \bibfield  {author} {\bibinfo {author} {\bibfnamefont {B.}~\bibnamefont {Derrida}}\ and\ \bibinfo {author} {\bibfnamefont {K.}~\bibnamefont {Mallick}},\ }\bibfield  {title} {\bibinfo {title} {Exact diffusion constant for the one-dimensional partially asymmetric exclusion model},\ }\href {https://doi.org/10.1088/0305-4470/30/4/007} {\bibfield  {journal} {\bibinfo  {journal} {Journal of Physics A: Mathematical and General}\ }\textbf {\bibinfo {volume} {30}},\ \bibinfo {pages} {1031} (\bibinfo {year} {1997})}\BibitemShut {NoStop}%
\bibitem [{\citenamefont {Berlioz}\ \emph {et~al.}(2025)\citenamefont {Berlioz}, \citenamefont {B\'enichou},\ and\ \citenamefont {Grabsch}}]{4j5q-j4ht}%
  \BibitemOpen
  \bibfield  {author} {\bibinfo {author} {\bibfnamefont {T.}~\bibnamefont {Berlioz}}, \bibinfo {author} {\bibfnamefont {O.}~\bibnamefont {B\'enichou}},\ and\ \bibinfo {author} {\bibfnamefont {A.}~\bibnamefont {Grabsch}},\ }\bibfield  {title} {\bibinfo {title} {Tracer and current fluctuations in driven diffusive systems},\ }\href {https://doi.org/10.1103/4j5q-j4ht} {\bibfield  {journal} {\bibinfo  {journal} {Phys. Rev. Lett.}\ }\textbf {\bibinfo {volume} {134}},\ \bibinfo {pages} {247101} (\bibinfo {year} {2025})}\BibitemShut {NoStop}%
\bibitem [{\citenamefont {Misakian}\ and\ \citenamefont {Kasianowicz}(2003)}]{Misakian:2003aa}%
  \BibitemOpen
  \bibfield  {author} {\bibinfo {author} {\bibfnamefont {M.}~\bibnamefont {Misakian}}\ and\ \bibinfo {author} {\bibfnamefont {J.~J.}\ \bibnamefont {Kasianowicz}},\ }\bibfield  {title} {\bibinfo {title} {Electrostatic influence on ion transport through the $\alpha$hl channel},\ }\href {https://doi.org/10.1007/s00232-003-0615-1} {\bibfield  {journal} {\bibinfo  {journal} {The Journal of Membrane Biology}\ }\textbf {\bibinfo {volume} {195}},\ \bibinfo {pages} {137} (\bibinfo {year} {2003})}\BibitemShut {NoStop}%
\bibitem [{\citenamefont {Nestorovich}\ \emph {et~al.}(2003)\citenamefont {Nestorovich}, \citenamefont {Rostovtseva},\ and\ \citenamefont {Bezrukov}}]{NESTOROVICH20033718}%
  \BibitemOpen
  \bibfield  {author} {\bibinfo {author} {\bibfnamefont {E.~M.}\ \bibnamefont {Nestorovich}}, \bibinfo {author} {\bibfnamefont {T.~K.}\ \bibnamefont {Rostovtseva}},\ and\ \bibinfo {author} {\bibfnamefont {S.~M.}\ \bibnamefont {Bezrukov}},\ }\bibfield  {title} {\bibinfo {title} {Residue ionization and ion transport through ompf channels},\ }\href {https://doi.org/https://doi.org/10.1016/S0006-3495(03)74788-2} {\bibfield  {journal} {\bibinfo  {journal} {Biophysical Journal}\ }\textbf {\bibinfo {volume} {85}},\ \bibinfo {pages} {3718} (\bibinfo {year} {2003})}\BibitemShut {NoStop}%
\bibitem [{\citenamefont {Noskov}\ \emph {et~al.}(2004)\citenamefont {Noskov}, \citenamefont {Im},\ and\ \citenamefont {Roux}}]{NOSKOV20042299}%
  \BibitemOpen
  \bibfield  {author} {\bibinfo {author} {\bibfnamefont {S.~Y.}\ \bibnamefont {Noskov}}, \bibinfo {author} {\bibfnamefont {W.}~\bibnamefont {Im}},\ and\ \bibinfo {author} {\bibfnamefont {B.}~\bibnamefont {Roux}},\ }\bibfield  {title} {\bibinfo {title} {Ion permeation through the $\alpha$-hemolysin channel: Theoretical studies based on brownian dynamics and poisson-nernst-plank electrodiffusion theory},\ }\href {https://doi.org/https://doi.org/10.1529/biophysj.104.044008} {\bibfield  {journal} {\bibinfo  {journal} {Biophysical Journal}\ }\textbf {\bibinfo {volume} {87}},\ \bibinfo {pages} {2299} (\bibinfo {year} {2004})}\BibitemShut {NoStop}%
\bibitem [{\citenamefont {Aksimentiev}\ and\ \citenamefont {Schulten}(2005)}]{AKSIMENTIEV20053745}%
  \BibitemOpen
  \bibfield  {author} {\bibinfo {author} {\bibfnamefont {A.}~\bibnamefont {Aksimentiev}}\ and\ \bibinfo {author} {\bibfnamefont {K.}~\bibnamefont {Schulten}},\ }\bibfield  {title} {\bibinfo {title} {Imaging $\alpha$-hemolysin with molecular dynamics: Ionic conductance, osmotic permeability, and the electrostatic potential map},\ }\href {https://doi.org/https://doi.org/10.1529/biophysj.104.058727} {\bibfield  {journal} {\bibinfo  {journal} {Biophysical Journal}\ }\textbf {\bibinfo {volume} {88}},\ \bibinfo {pages} {3745} (\bibinfo {year} {2005})}\BibitemShut {NoStop}%
\bibitem [{\citenamefont {Li}\ \emph {et~al.}(2007)\citenamefont {Li}, \citenamefont {Berke}, \citenamefont {Chen},\ and\ \citenamefont {Jiang}}]{10.1085/200609655}%
  \BibitemOpen
  \bibfield  {author} {\bibinfo {author} {\bibfnamefont {Y.}~\bibnamefont {Li}}, \bibinfo {author} {\bibfnamefont {I.}~\bibnamefont {Berke}}, \bibinfo {author} {\bibfnamefont {L.}~\bibnamefont {Chen}},\ and\ \bibinfo {author} {\bibfnamefont {Y.}~\bibnamefont {Jiang}},\ }\bibfield  {title} {\bibinfo {title} {Gating and inward rectifying properties of the mthk k+ channel with and without the gating ring},\ }\href {https://doi.org/10.1085/jgp.200609655} {\bibfield  {journal} {\bibinfo  {journal} {Journal of General Physiology}\ }\textbf {\bibinfo {volume} {129}},\ \bibinfo {pages} {109} (\bibinfo {year} {2007})}\BibitemShut {NoStop}%
\bibitem [{\citenamefont {Bhattacharya}\ \emph {et~al.}(2011)\citenamefont {Bhattacharya}, \citenamefont {Muzard}, \citenamefont {Payet}, \citenamefont {Math{\'e}}, \citenamefont {Bockelmann}, \citenamefont {Aksimentiev},\ and\ \citenamefont {Viasnoff}}]{Bhattacharya:2011aa}%
  \BibitemOpen
  \bibfield  {author} {\bibinfo {author} {\bibfnamefont {S.}~\bibnamefont {Bhattacharya}}, \bibinfo {author} {\bibfnamefont {J.}~\bibnamefont {Muzard}}, \bibinfo {author} {\bibfnamefont {L.}~\bibnamefont {Payet}}, \bibinfo {author} {\bibfnamefont {J.}~\bibnamefont {Math{\'e}}}, \bibinfo {author} {\bibfnamefont {U.}~\bibnamefont {Bockelmann}}, \bibinfo {author} {\bibfnamefont {A.}~\bibnamefont {Aksimentiev}},\ and\ \bibinfo {author} {\bibfnamefont {V.}~\bibnamefont {Viasnoff}},\ }\bibfield  {title} {\bibinfo {title} {Rectification of the current in $\alpha$-hemolysin pore depends on the cation type: The alkali series probed by molecular dynamics simulations and experiments},\ }\href {https://doi.org/10.1021/jp111441p} {\bibfield  {journal} {\bibinfo  {journal} {The Journal of Physical Chemistry C}\ }\textbf {\bibinfo {volume} {115}},\ \bibinfo {pages} {4255} (\bibinfo {year} {2011})}\BibitemShut {NoStop}%
\bibitem [{\citenamefont {Piguet}\ \emph {et~al.}(2014)\citenamefont {Piguet}, \citenamefont {Discala}, \citenamefont {Breton}, \citenamefont {Pelta}, \citenamefont {Bacri},\ and\ \citenamefont {Oukhaled}}]{Piguet:2014aa}%
  \BibitemOpen
  \bibfield  {author} {\bibinfo {author} {\bibfnamefont {F.}~\bibnamefont {Piguet}}, \bibinfo {author} {\bibfnamefont {F.}~\bibnamefont {Discala}}, \bibinfo {author} {\bibfnamefont {M.-F.}\ \bibnamefont {Breton}}, \bibinfo {author} {\bibfnamefont {J.}~\bibnamefont {Pelta}}, \bibinfo {author} {\bibfnamefont {L.}~\bibnamefont {Bacri}},\ and\ \bibinfo {author} {\bibfnamefont {A.}~\bibnamefont {Oukhaled}},\ }\bibfield  {title} {\bibinfo {title} {Electroosmosis through $\alpha$-hemolysin that depends on alkali cation type},\ }\href {https://doi.org/10.1021/jz502360c} {\bibfield  {journal} {\bibinfo  {journal} {The Journal of Physical Chemistry Letters}\ }\textbf {\bibinfo {volume} {5}},\ \bibinfo {pages} {4362} (\bibinfo {year} {2014})}\BibitemShut {NoStop}%
\bibitem [{\citenamefont {Payet}\ \emph {et~al.}(2015)\citenamefont {Payet}, \citenamefont {Martinho}, \citenamefont {Merstorf}, \citenamefont {Pastoriza-Gallego}, \citenamefont {Pelta}, \citenamefont {Viasnoff}, \citenamefont {Auvray}, \citenamefont {Muthukumar},\ and\ \citenamefont {Math{\'e}}}]{PAYET20151600}%
  \BibitemOpen
  \bibfield  {author} {\bibinfo {author} {\bibfnamefont {L.}~\bibnamefont {Payet}}, \bibinfo {author} {\bibfnamefont {M.}~\bibnamefont {Martinho}}, \bibinfo {author} {\bibfnamefont {C.}~\bibnamefont {Merstorf}}, \bibinfo {author} {\bibfnamefont {M.}~\bibnamefont {Pastoriza-Gallego}}, \bibinfo {author} {\bibfnamefont {J.}~\bibnamefont {Pelta}}, \bibinfo {author} {\bibfnamefont {V.}~\bibnamefont {Viasnoff}}, \bibinfo {author} {\bibfnamefont {L.}~\bibnamefont {Auvray}}, \bibinfo {author} {\bibfnamefont {M.}~\bibnamefont {Muthukumar}},\ and\ \bibinfo {author} {\bibfnamefont {J.}~\bibnamefont {Math{\'e}}},\ }\bibfield  {title} {\bibinfo {title} {Temperature effect on ionic current and ssdna transport through nanopores},\ }\href {https://doi.org/https://doi.org/10.1016/j.bpj.2015.08.043} {\bibfield  {journal} {\bibinfo  {journal} {Biophysical Journal}\ }\textbf {\bibinfo {volume} {109}},\ \bibinfo {pages} {1600} (\bibinfo {year} {2015})}\BibitemShut {NoStop}%
\bibitem [{\citenamefont {Manara}\ \emph {et~al.}(2015)\citenamefont {Manara}, \citenamefont {Guy}, \citenamefont {Wallace},\ and\ \citenamefont {Khalid}}]{Manara:2015aa}%
  \BibitemOpen
  \bibfield  {author} {\bibinfo {author} {\bibfnamefont {R.~M.~A.}\ \bibnamefont {Manara}}, \bibinfo {author} {\bibfnamefont {A.~T.}\ \bibnamefont {Guy}}, \bibinfo {author} {\bibfnamefont {E.~J.}\ \bibnamefont {Wallace}},\ and\ \bibinfo {author} {\bibfnamefont {S.}~\bibnamefont {Khalid}},\ }\bibfield  {title} {\bibinfo {title} {Free-energy calculations reveal the subtle differences in the interactions of dna bases with $\alpha$-hemolysin},\ }\href {https://doi.org/10.1021/ct501081h} {\bibfield  {journal} {\bibinfo  {journal} {Journal of Chemical Theory and Computation}\ }\textbf {\bibinfo {volume} {11}},\ \bibinfo {pages} {810} (\bibinfo {year} {2015})}\BibitemShut {NoStop}%
\bibitem [{\citenamefont {Ionescu}\ \emph {et~al.}(2017)\citenamefont {Ionescu}, \citenamefont {Lee}, \citenamefont {Housden}, \citenamefont {Kaminska}, \citenamefont {Kleanthous},\ and\ \citenamefont {Bayley}}]{https://doi.org/10.1002/cbic.201600644}%
  \BibitemOpen
  \bibfield  {author} {\bibinfo {author} {\bibfnamefont {S.~A.}\ \bibnamefont {Ionescu}}, \bibinfo {author} {\bibfnamefont {S.}~\bibnamefont {Lee}}, \bibinfo {author} {\bibfnamefont {N.~G.}\ \bibnamefont {Housden}}, \bibinfo {author} {\bibfnamefont {R.}~\bibnamefont {Kaminska}}, \bibinfo {author} {\bibfnamefont {C.}~\bibnamefont {Kleanthous}},\ and\ \bibinfo {author} {\bibfnamefont {H.}~\bibnamefont {Bayley}},\ }\bibfield  {title} {\bibinfo {title} {Orientation of the ompf porin in planar lipid bilayers},\ }\href {https://doi.org/https://doi.org/10.1002/cbic.201600644} {\bibfield  {journal} {\bibinfo  {journal} {ChemBioChem}\ }\textbf {\bibinfo {volume} {18}},\ \bibinfo {pages} {554} (\bibinfo {year} {2017})}\BibitemShut {NoStop}%
\bibitem [{\citenamefont {Zhou}\ \emph {et~al.}(2020)\citenamefont {Zhou}, \citenamefont {Qiu}, \citenamefont {Guo},\ and\ \citenamefont {Guo}}]{Zhou:2020aa}%
  \BibitemOpen
  \bibfield  {author} {\bibinfo {author} {\bibfnamefont {W.}~\bibnamefont {Zhou}}, \bibinfo {author} {\bibfnamefont {H.}~\bibnamefont {Qiu}}, \bibinfo {author} {\bibfnamefont {Y.}~\bibnamefont {Guo}},\ and\ \bibinfo {author} {\bibfnamefont {W.}~\bibnamefont {Guo}},\ }\bibfield  {title} {\bibinfo {title} {Molecular insights into distinct detection properties of $\alpha$-hemolysin, mspa, csgg, and aerolysin nanopore sensors},\ }\href {https://doi.org/10.1021/acs.jpcb.9b10702} {\bibfield  {journal} {\bibinfo  {journal} {The Journal of Physical Chemistry B}\ }\textbf {\bibinfo {volume} {124}},\ \bibinfo {pages} {1611} (\bibinfo {year} {2020})}\BibitemShut {NoStop}%
\bibitem [{\citenamefont {Dessaux}\ \emph {et~al.}(2022)\citenamefont {Dessaux}, \citenamefont {Math{\'e}}, \citenamefont {Ramirez},\ and\ \citenamefont {Basdevant}}]{Dessaux:2022aa}%
  \BibitemOpen
  \bibfield  {author} {\bibinfo {author} {\bibfnamefont {D.}~\bibnamefont {Dessaux}}, \bibinfo {author} {\bibfnamefont {J.}~\bibnamefont {Math{\'e}}}, \bibinfo {author} {\bibfnamefont {R.}~\bibnamefont {Ramirez}},\ and\ \bibinfo {author} {\bibfnamefont {N.}~\bibnamefont {Basdevant}},\ }\bibfield  {title} {\bibinfo {title} {Current rectification and ionic selectivity of $\alpha$-hemolysin: Coarse-grained molecular dynamics simulations},\ }\href {https://doi.org/10.1021/acs.jpcb.2c01028} {\bibfield  {journal} {\bibinfo  {journal} {The Journal of Physical Chemistry B}\ }\textbf {\bibinfo {volume} {126}},\ \bibinfo {pages} {4189} (\bibinfo {year} {2022})}\BibitemShut {NoStop}%
\bibitem [{\citenamefont {Donoghue}\ \emph {et~al.}(2023)\citenamefont {Donoghue}, \citenamefont {Winterhalter},\ and\ \citenamefont {Gutsmann}}]{membranes13050517}%
  \BibitemOpen
  \bibfield  {author} {\bibinfo {author} {\bibfnamefont {A.}~\bibnamefont {Donoghue}}, \bibinfo {author} {\bibfnamefont {M.}~\bibnamefont {Winterhalter}},\ and\ \bibinfo {author} {\bibfnamefont {T.}~\bibnamefont {Gutsmann}},\ }\bibfield  {title} {\bibinfo {title} {Influence of membrane asymmetry on ompf insertion, orientation and function},\ }\bibfield  {journal} {\bibinfo  {journal} {Membranes}\ }\textbf {\bibinfo {volume} {13}},\ \href {https://doi.org/10.3390/membranes13050517} {10.3390/membranes13050517} (\bibinfo {year} {2023})\BibitemShut {NoStop}%
\bibitem [{\citenamefont {Siwy}(2006)}]{https://doi.org/10.1002/adfm.200500471}%
  \BibitemOpen
  \bibfield  {author} {\bibinfo {author} {\bibfnamefont {Z.}~\bibnamefont {Siwy}},\ }\bibfield  {title} {\bibinfo {title} {Ion-current rectification in nanopores and nanotubes with broken symmetry},\ }\href {https://doi.org/https://doi.org/10.1002/adfm.200500471} {\bibfield  {journal} {\bibinfo  {journal} {Advanced Functional Materials}\ }\textbf {\bibinfo {volume} {16}},\ \bibinfo {pages} {735} (\bibinfo {year} {2006})}\BibitemShut {NoStop}%
\bibitem [{\citenamefont {Yan}\ \emph {et~al.}(2013)\citenamefont {Yan}, \citenamefont {Wang}, \citenamefont {Xue},\ and\ \citenamefont {Chang}}]{10.1063/1.4776216}%
  \BibitemOpen
  \bibfield  {author} {\bibinfo {author} {\bibfnamefont {Y.}~\bibnamefont {Yan}}, \bibinfo {author} {\bibfnamefont {L.}~\bibnamefont {Wang}}, \bibinfo {author} {\bibfnamefont {J.}~\bibnamefont {Xue}},\ and\ \bibinfo {author} {\bibfnamefont {H.-C.}\ \bibnamefont {Chang}},\ }\bibfield  {title} {\bibinfo {title} {Ion current rectification inversion in conic nanopores: Nonequilibrium ion transport biased by ion selectivity and spatial asymmetry},\ }\href {https://doi.org/10.1063/1.4776216} {\bibfield  {journal} {\bibinfo  {journal} {The Journal of Chemical Physics}\ }\textbf {\bibinfo {volume} {138}},\ \bibinfo {pages} {044706} (\bibinfo {year} {2013})}\BibitemShut {NoStop}%
\bibitem [{\citenamefont {Gamble}\ \emph {et~al.}(2014)\citenamefont {Gamble}, \citenamefont {Decker}, \citenamefont {Plett}, \citenamefont {Pevarnik}, \citenamefont {Pietschmann}, \citenamefont {Vlassiouk}, \citenamefont {Aksimentiev},\ and\ \citenamefont {Siwy}}]{Gamble:2014aa}%
  \BibitemOpen
  \bibfield  {author} {\bibinfo {author} {\bibfnamefont {T.}~\bibnamefont {Gamble}}, \bibinfo {author} {\bibfnamefont {K.}~\bibnamefont {Decker}}, \bibinfo {author} {\bibfnamefont {T.~S.}\ \bibnamefont {Plett}}, \bibinfo {author} {\bibfnamefont {M.}~\bibnamefont {Pevarnik}}, \bibinfo {author} {\bibfnamefont {J.-F.}\ \bibnamefont {Pietschmann}}, \bibinfo {author} {\bibfnamefont {I.}~\bibnamefont {Vlassiouk}}, \bibinfo {author} {\bibfnamefont {A.}~\bibnamefont {Aksimentiev}},\ and\ \bibinfo {author} {\bibfnamefont {Z.~S.}\ \bibnamefont {Siwy}},\ }\bibfield  {title} {\bibinfo {title} {Rectification of ion current in nanopores depends on the type of monovalent cations: Experiments and modeling},\ }\href {https://doi.org/10.1021/jp501492g} {\bibfield  {journal} {\bibinfo  {journal} {The Journal of Physical Chemistry C}\ }\textbf {\bibinfo {volume} {118}},\ \bibinfo {pages} {9809} (\bibinfo {year} {2014})}\BibitemShut {NoStop}%
\bibitem [{\citenamefont {Su}\ \emph {et~al.}(2023)\citenamefont {Su}, \citenamefont {Hung}, \citenamefont {Fauziah}, \citenamefont {Siwy},\ and\ \citenamefont {Yeh}}]{SU2023141064}%
  \BibitemOpen
  \bibfield  {author} {\bibinfo {author} {\bibfnamefont {Y.-S.}\ \bibnamefont {Su}}, \bibinfo {author} {\bibfnamefont {W.-H.}\ \bibnamefont {Hung}}, \bibinfo {author} {\bibfnamefont {A.~R.}\ \bibnamefont {Fauziah}}, \bibinfo {author} {\bibfnamefont {Z.~S.}\ \bibnamefont {Siwy}},\ and\ \bibinfo {author} {\bibfnamefont {L.-H.}\ \bibnamefont {Yeh}},\ }\bibfield  {title} {\bibinfo {title} {A ph gradient induced rectification inversion in asymmetric nanochannels leads to remarkably improved osmotic power},\ }\href {https://doi.org/https://doi.org/10.1016/j.cej.2022.141064} {\bibfield  {journal} {\bibinfo  {journal} {Chemical Engineering Journal}\ }\textbf {\bibinfo {volume} {456}},\ \bibinfo {pages} {141064} (\bibinfo {year} {2023})}\BibitemShut {NoStop}%
\bibitem [{\citenamefont {Tripathy}\ and\ \citenamefont {Barma}(1997)}]{PhysRevLett.78.3039}%
  \BibitemOpen
  \bibfield  {author} {\bibinfo {author} {\bibfnamefont {G.}~\bibnamefont {Tripathy}}\ and\ \bibinfo {author} {\bibfnamefont {M.}~\bibnamefont {Barma}},\ }\bibfield  {title} {\bibinfo {title} {Steady state and dynamics of driven diffusive systems with quenched disorder},\ }\href {https://doi.org/10.1103/PhysRevLett.78.3039} {\bibfield  {journal} {\bibinfo  {journal} {Phys. Rev. Lett.}\ }\textbf {\bibinfo {volume} {78}},\ \bibinfo {pages} {3039} (\bibinfo {year} {1997})}\BibitemShut {NoStop}%
\bibitem [{\citenamefont {Tripathy}\ and\ \citenamefont {Barma}(1998)}]{PhysRevE.58.1911}%
  \BibitemOpen
  \bibfield  {author} {\bibinfo {author} {\bibfnamefont {G.}~\bibnamefont {Tripathy}}\ and\ \bibinfo {author} {\bibfnamefont {M.}~\bibnamefont {Barma}},\ }\bibfield  {title} {\bibinfo {title} {Driven lattice gases with quenched disorder: Exact results and different macroscopic regimes},\ }\href {https://doi.org/10.1103/PhysRevE.58.1911} {\bibfield  {journal} {\bibinfo  {journal} {Phys. Rev. E}\ }\textbf {\bibinfo {volume} {58}},\ \bibinfo {pages} {1911} (\bibinfo {year} {1998})}\BibitemShut {NoStop}%
\bibitem [{\citenamefont {Goldstein}\ and\ \citenamefont {Speer}(1998)}]{PhysRevE.58.4226}%
  \BibitemOpen
  \bibfield  {author} {\bibinfo {author} {\bibfnamefont {S.}~\bibnamefont {Goldstein}}\ and\ \bibinfo {author} {\bibfnamefont {E.~R.}\ \bibnamefont {Speer}},\ }\bibfield  {title} {\bibinfo {title} {Reflection invariance of the current in the totally asymmetric simple exclusion process with disorder},\ }\href {https://doi.org/10.1103/PhysRevE.58.4226} {\bibfield  {journal} {\bibinfo  {journal} {Phys. Rev. E}\ }\textbf {\bibinfo {volume} {58}},\ \bibinfo {pages} {4226} (\bibinfo {year} {1998})}\BibitemShut {NoStop}%
\bibitem [{\citenamefont {Enaud}\ and\ \citenamefont {Derrida}(2004)}]{Enaud:2004aa}%
  \BibitemOpen
  \bibfield  {author} {\bibinfo {author} {\bibfnamefont {C.}~\bibnamefont {Enaud}}\ and\ \bibinfo {author} {\bibfnamefont {B.}~\bibnamefont {Derrida}},\ }\bibfield  {title} {\bibinfo {title} {Sample-dependent phase transitions in disordered exclusion models},\ }\href {https://doi.org/10.1209/epl/i2003-10153-8} {\bibfield  {journal} {\bibinfo  {journal} {Europhysics Letters}\ }\textbf {\bibinfo {volume} {66}},\ \bibinfo {pages} {83} (\bibinfo {year} {2004})}\BibitemShut {NoStop}%
\bibitem [{\citenamefont {Harris}\ and\ \citenamefont {Stinchcombe}(2004)}]{PhysRevE.70.016108}%
  \BibitemOpen
  \bibfield  {author} {\bibinfo {author} {\bibfnamefont {R.~J.}\ \bibnamefont {Harris}}\ and\ \bibinfo {author} {\bibfnamefont {R.~B.}\ \bibnamefont {Stinchcombe}},\ }\bibfield  {title} {\bibinfo {title} {Disordered asymmetric simple exclusion process: Mean-field treatment},\ }\href {https://doi.org/10.1103/PhysRevE.70.016108} {\bibfield  {journal} {\bibinfo  {journal} {Phys. Rev. E}\ }\textbf {\bibinfo {volume} {70}},\ \bibinfo {pages} {016108} (\bibinfo {year} {2004})}\BibitemShut {NoStop}%
\bibitem [{\citenamefont {Juh\'asz}\ \emph {et~al.}(2005)\citenamefont {Juh\'asz}, \citenamefont {Santen},\ and\ \citenamefont {Igl\'oi}}]{PhysRevLett.94.010601}%
  \BibitemOpen
  \bibfield  {author} {\bibinfo {author} {\bibfnamefont {R.}~\bibnamefont {Juh\'asz}}, \bibinfo {author} {\bibfnamefont {L.}~\bibnamefont {Santen}},\ and\ \bibinfo {author} {\bibfnamefont {F.}~\bibnamefont {Igl\'oi}},\ }\bibfield  {title} {\bibinfo {title} {Partially asymmetric exclusion models with quenched disorder},\ }\href {https://doi.org/10.1103/PhysRevLett.94.010601} {\bibfield  {journal} {\bibinfo  {journal} {Phys. Rev. Lett.}\ }\textbf {\bibinfo {volume} {94}},\ \bibinfo {pages} {010601} (\bibinfo {year} {2005})}\BibitemShut {NoStop}%
\bibitem [{\citenamefont {Juh\'asz}\ \emph {et~al.}(2006)\citenamefont {Juh\'asz}, \citenamefont {Santen},\ and\ \citenamefont {Igl\'oi}}]{PhysRevE.74.061101}%
  \BibitemOpen
  \bibfield  {author} {\bibinfo {author} {\bibfnamefont {R.}~\bibnamefont {Juh\'asz}}, \bibinfo {author} {\bibfnamefont {L.}~\bibnamefont {Santen}},\ and\ \bibinfo {author} {\bibfnamefont {F.}~\bibnamefont {Igl\'oi}},\ }\bibfield  {title} {\bibinfo {title} {Partially asymmetric exclusion processes with sitewise disorder},\ }\href {https://doi.org/10.1103/PhysRevE.74.061101} {\bibfield  {journal} {\bibinfo  {journal} {Phys. Rev. E}\ }\textbf {\bibinfo {volume} {74}},\ \bibinfo {pages} {061101} (\bibinfo {year} {2006})}\BibitemShut {NoStop}%
\bibitem [{\citenamefont {Ebrahim~Foulaadvand}\ \emph {et~al.}(2007)\citenamefont {Ebrahim~Foulaadvand}, \citenamefont {Chaaboki},\ and\ \citenamefont {Saalehi}}]{PhysRevE.75.011127}%
  \BibitemOpen
  \bibfield  {author} {\bibinfo {author} {\bibfnamefont {M.}~\bibnamefont {Ebrahim~Foulaadvand}}, \bibinfo {author} {\bibfnamefont {S.}~\bibnamefont {Chaaboki}},\ and\ \bibinfo {author} {\bibfnamefont {M.}~\bibnamefont {Saalehi}},\ }\bibfield  {title} {\bibinfo {title} {Characteristics of the asymmetric simple exclusion process in the presence of quenched spatial disorder},\ }\href {https://doi.org/10.1103/PhysRevE.75.011127} {\bibfield  {journal} {\bibinfo  {journal} {Phys. Rev. E}\ }\textbf {\bibinfo {volume} {75}},\ \bibinfo {pages} {011127} (\bibinfo {year} {2007})}\BibitemShut {NoStop}%
\bibitem [{\citenamefont {Greulich}\ and\ \citenamefont {Schadschneider}(2008)}]{Greulich:2008aa}%
  \BibitemOpen
  \bibfield  {author} {\bibinfo {author} {\bibfnamefont {P.}~\bibnamefont {Greulich}}\ and\ \bibinfo {author} {\bibfnamefont {A.}~\bibnamefont {Schadschneider}},\ }\bibfield  {title} {\bibinfo {title} {Single-bottleneck approximation for driven lattice gases with disorder and open boundary conditions},\ }\href {https://doi.org/10.1088/1742-5468/2008/04/P04009} {\bibfield  {journal} {\bibinfo  {journal} {Journal of Statistical Mechanics: Theory and Experiment}\ }\textbf {\bibinfo {volume} {2008}},\ \bibinfo {pages} {P04009} (\bibinfo {year} {2008})}\BibitemShut {NoStop}%
\bibitem [{\citenamefont {Foulaadvand}\ \emph {et~al.}(2008)\citenamefont {Foulaadvand}, \citenamefont {Kolomeisky},\ and\ \citenamefont {Teymouri}}]{PhysRevE.78.061116}%
  \BibitemOpen
  \bibfield  {author} {\bibinfo {author} {\bibfnamefont {M.~E.}\ \bibnamefont {Foulaadvand}}, \bibinfo {author} {\bibfnamefont {A.~B.}\ \bibnamefont {Kolomeisky}},\ and\ \bibinfo {author} {\bibfnamefont {H.}~\bibnamefont {Teymouri}},\ }\bibfield  {title} {\bibinfo {title} {Asymmetric exclusion processes with disorder: Effect of correlations},\ }\href {https://doi.org/10.1103/PhysRevE.78.061116} {\bibfield  {journal} {\bibinfo  {journal} {Phys. Rev. E}\ }\textbf {\bibinfo {volume} {78}},\ \bibinfo {pages} {061116} (\bibinfo {year} {2008})}\BibitemShut {NoStop}%
\bibitem [{\citenamefont {Concannon}\ and\ \citenamefont {Blythe}(2014)}]{PhysRevLett.112.050603}%
  \BibitemOpen
  \bibfield  {author} {\bibinfo {author} {\bibfnamefont {R.~J.}\ \bibnamefont {Concannon}}\ and\ \bibinfo {author} {\bibfnamefont {R.~A.}\ \bibnamefont {Blythe}},\ }\bibfield  {title} {\bibinfo {title} {Spatiotemporally complete condensation in a non-poissonian exclusion process},\ }\href {https://doi.org/10.1103/PhysRevLett.112.050603} {\bibfield  {journal} {\bibinfo  {journal} {Phys. Rev. Lett.}\ }\textbf {\bibinfo {volume} {112}},\ \bibinfo {pages} {050603} (\bibinfo {year} {2014})}\BibitemShut {NoStop}%
\bibitem [{\citenamefont {Bahadoran}\ and\ \citenamefont {Bodineau}(2015)}]{10.1214/14-BJPS277}%
  \BibitemOpen
  \bibfield  {author} {\bibinfo {author} {\bibfnamefont {C.}~\bibnamefont {Bahadoran}}\ and\ \bibinfo {author} {\bibfnamefont {T.}~\bibnamefont {Bodineau}},\ }\bibfield  {title} {\bibinfo {title} {{Properties and conjectures for the flux of TASEP with site disorder}},\ }\href {https://doi.org/10.1214/14-BJPS277} {\bibfield  {journal} {\bibinfo  {journal} {Braz. J. Probab. Stat.}\ }\textbf {\bibinfo {volume} {29}},\ \bibinfo {pages} {282 } (\bibinfo {year} {2015})}\BibitemShut {NoStop}%
\bibitem [{\citenamefont {Banerjee}\ and\ \citenamefont {Basu}(2020)}]{PhysRevResearch.2.013025}%
  \BibitemOpen
  \bibfield  {author} {\bibinfo {author} {\bibfnamefont {T.}~\bibnamefont {Banerjee}}\ and\ \bibinfo {author} {\bibfnamefont {A.}~\bibnamefont {Basu}},\ }\bibfield  {title} {\bibinfo {title} {Smooth or shock: Universality in closed inhomogeneous driven single file motions},\ }\href {https://doi.org/10.1103/PhysRevResearch.2.013025} {\bibfield  {journal} {\bibinfo  {journal} {Phys. Rev. Res.}\ }\textbf {\bibinfo {volume} {2}},\ \bibinfo {pages} {013025} (\bibinfo {year} {2020})}\BibitemShut {NoStop}%
\bibitem [{\citenamefont {Haldar}\ and\ \citenamefont {Basu}(2020)}]{PhysRevResearch.2.043073}%
  \BibitemOpen
  \bibfield  {author} {\bibinfo {author} {\bibfnamefont {A.}~\bibnamefont {Haldar}}\ and\ \bibinfo {author} {\bibfnamefont {A.}~\bibnamefont {Basu}},\ }\bibfield  {title} {\bibinfo {title} {Marching on a rugged landscape: Universality in disordered asymmetric exclusion processes},\ }\href {https://doi.org/10.1103/PhysRevResearch.2.043073} {\bibfield  {journal} {\bibinfo  {journal} {Phys. Rev. Res.}\ }\textbf {\bibinfo {volume} {2}},\ \bibinfo {pages} {043073} (\bibinfo {year} {2020})}\BibitemShut {NoStop}%
\bibitem [{\citenamefont {Goswami}\ \emph {et~al.}(2022)\citenamefont {Goswami}, \citenamefont {Chatterjee},\ and\ \citenamefont {Mukherjee}}]{Goswami:2022aa}%
  \BibitemOpen
  \bibfield  {author} {\bibinfo {author} {\bibfnamefont {A.}~\bibnamefont {Goswami}}, \bibinfo {author} {\bibfnamefont {M.}~\bibnamefont {Chatterjee}},\ and\ \bibinfo {author} {\bibfnamefont {S.}~\bibnamefont {Mukherjee}},\ }\bibfield  {title} {\bibinfo {title} {Steady states and phase transitions in heterogeneous asymmetric exclusion processes},\ }\href {https://doi.org/10.1088/1742-5468/aca2a0} {\bibfield  {journal} {\bibinfo  {journal} {Journal of Statistical Mechanics: Theory and Experiment}\ }\textbf {\bibinfo {volume} {2022}},\ \bibinfo {pages} {123209} (\bibinfo {year} {2022})}\BibitemShut {NoStop}%
\bibitem [{\citenamefont {Sakai}\ and\ \citenamefont {Akimoto}(2023{\natexlab{a}})}]{PhysRevE.107.L052103}%
  \BibitemOpen
  \bibfield  {author} {\bibinfo {author} {\bibfnamefont {I.}~\bibnamefont {Sakai}}\ and\ \bibinfo {author} {\bibfnamefont {T.}~\bibnamefont {Akimoto}},\ }\bibfield  {title} {\bibinfo {title} {Non-self-averaging of current in a totally asymmetric simple exclusion process with quenched disorder},\ }\href {https://doi.org/10.1103/PhysRevE.107.L052103} {\bibfield  {journal} {\bibinfo  {journal} {Phys. Rev. E}\ }\textbf {\bibinfo {volume} {107}},\ \bibinfo {pages} {L052103} (\bibinfo {year} {2023}{\natexlab{a}})}\BibitemShut {NoStop}%
\bibitem [{\citenamefont {Sakai}\ and\ \citenamefont {Akimoto}(2023{\natexlab{b}})}]{PhysRevE.107.054131}%
  \BibitemOpen
  \bibfield  {author} {\bibinfo {author} {\bibfnamefont {I.}~\bibnamefont {Sakai}}\ and\ \bibinfo {author} {\bibfnamefont {T.}~\bibnamefont {Akimoto}},\ }\bibfield  {title} {\bibinfo {title} {Sample-to-sample fluctuations of transport coefficients in the totally asymmetric simple exclusion process with quenched disorder},\ }\href {https://doi.org/10.1103/PhysRevE.107.054131} {\bibfield  {journal} {\bibinfo  {journal} {Phys. Rev. E}\ }\textbf {\bibinfo {volume} {107}},\ \bibinfo {pages} {054131} (\bibinfo {year} {2023}{\natexlab{b}})}\BibitemShut {NoStop}%
\bibitem [{\citenamefont {Bouchaud}\ and\ \citenamefont {Georges}(1990)}]{BouchaudGeorfes}%
  \BibitemOpen
  \bibfield  {author} {\bibinfo {author} {\bibfnamefont {J.~P.}\ \bibnamefont {Bouchaud}}\ and\ \bibinfo {author} {\bibfnamefont {A.}~\bibnamefont {Georges}},\ }\bibfield  {title} {\bibinfo {title} {Anomalous diffusion in disordered media: Statistical mechanisms, models and physical applications},\ }\href@noop {} {\bibfield  {journal} {\bibinfo  {journal} {Phys. Rep.}\ }\textbf {\bibinfo {volume} {195}} (\bibinfo {year} {1990})}\BibitemShut {NoStop}%
\bibitem [{\citenamefont {Hanney}\ and\ \citenamefont {Evans}(2003)}]{Hanney:2003aa}%
  \BibitemOpen
  \bibfield  {author} {\bibinfo {author} {\bibfnamefont {T.}~\bibnamefont {Hanney}}\ and\ \bibinfo {author} {\bibfnamefont {M.~R.}\ \bibnamefont {Evans}},\ }\bibfield  {title} {\bibinfo {title} {Einstein relation for nonequilibrium steady states},\ }\href {https://doi.org/10.1023/A:1023068619793} {\bibfield  {journal} {\bibinfo  {journal} {Journal of Statistical Physics}\ }\textbf {\bibinfo {volume} {111}},\ \bibinfo {pages} {1377} (\bibinfo {year} {2003})}\BibitemShut {NoStop}%
\bibitem [{\citenamefont {Sakai}\ and\ \citenamefont {Akimoto}(2025)}]{PhysRevE.111.014134}%
  \BibitemOpen
  \bibfield  {author} {\bibinfo {author} {\bibfnamefont {I.}~\bibnamefont {Sakai}}\ and\ \bibinfo {author} {\bibfnamefont {T.}~\bibnamefont {Akimoto}},\ }\bibfield  {title} {\bibinfo {title} {Unexpected effects of disorder on current fluctuations in the symmetric simple exclusion process},\ }\href {https://doi.org/10.1103/PhysRevE.111.014134} {\bibfield  {journal} {\bibinfo  {journal} {Phys. Rev. E}\ }\textbf {\bibinfo {volume} {111}},\ \bibinfo {pages} {014134} (\bibinfo {year} {2025})}\BibitemShut {NoStop}%
\bibitem [{\citenamefont {Akimoto}\ and\ \citenamefont {Saito}(2020)}]{PhysRevE.101.042133}%
  \BibitemOpen
  \bibfield  {author} {\bibinfo {author} {\bibfnamefont {T.}~\bibnamefont {Akimoto}}\ and\ \bibinfo {author} {\bibfnamefont {K.}~\bibnamefont {Saito}},\ }\bibfield  {title} {\bibinfo {title} {Trace of anomalous diffusion in a biased quenched trap model},\ }\href {https://doi.org/10.1103/PhysRevE.101.042133} {\bibfield  {journal} {\bibinfo  {journal} {Phys. Rev. E}\ }\textbf {\bibinfo {volume} {101}},\ \bibinfo {pages} {042133} (\bibinfo {year} {2020})}\BibitemShut {NoStop}%
\bibitem [{\citenamefont {Hanes}\ \emph {et~al.}(2012)\citenamefont {Hanes}, \citenamefont {Dalle-Ferrier}, \citenamefont {Schmiedeberg}, \citenamefont {Jenkins},\ and\ \citenamefont {Egelhaaf}}]{C2SM07102A}%
  \BibitemOpen
  \bibfield  {author} {\bibinfo {author} {\bibfnamefont {R.~D.~L.}\ \bibnamefont {Hanes}}, \bibinfo {author} {\bibfnamefont {C.}~\bibnamefont {Dalle-Ferrier}}, \bibinfo {author} {\bibfnamefont {M.}~\bibnamefont {Schmiedeberg}}, \bibinfo {author} {\bibfnamefont {M.~C.}\ \bibnamefont {Jenkins}},\ and\ \bibinfo {author} {\bibfnamefont {S.~U.}\ \bibnamefont {Egelhaaf}},\ }\bibfield  {title} {\bibinfo {title} {Colloids in one dimensional random energy landscapes},\ }\href {https://doi.org/10.1039/C2SM07102A} {\bibfield  {journal} {\bibinfo  {journal} {Soft Matter}\ }\textbf {\bibinfo {volume} {8}},\ \bibinfo {pages} {2714} (\bibinfo {year} {2012})}\BibitemShut {NoStop}%
\bibitem [{\citenamefont {Yang}\ \emph {et~al.}(2010)\citenamefont {Yang}, \citenamefont {Yang}, \citenamefont {Kim}, \citenamefont {Jeon}, \citenamefont {Oh}, \citenamefont {Choi}, \citenamefont {Hahn},\ and\ \citenamefont {Kim}}]{Yang:2010aa}%
  \BibitemOpen
  \bibfield  {author} {\bibinfo {author} {\bibfnamefont {S.~Y.}\ \bibnamefont {Yang}}, \bibinfo {author} {\bibfnamefont {J.-A.}\ \bibnamefont {Yang}}, \bibinfo {author} {\bibfnamefont {E.-S.}\ \bibnamefont {Kim}}, \bibinfo {author} {\bibfnamefont {G.}~\bibnamefont {Jeon}}, \bibinfo {author} {\bibfnamefont {E.~J.}\ \bibnamefont {Oh}}, \bibinfo {author} {\bibfnamefont {K.~Y.}\ \bibnamefont {Choi}}, \bibinfo {author} {\bibfnamefont {S.~K.}\ \bibnamefont {Hahn}},\ and\ \bibinfo {author} {\bibfnamefont {J.~K.}\ \bibnamefont {Kim}},\ }\bibfield  {title} {\bibinfo {title} {Single-file diffusion of protein drugs through cylindrical nanochannels},\ }\href {https://doi.org/10.1021/nn100464u} {\bibfield  {journal} {\bibinfo  {journal} {ACS Nano}\ }\textbf {\bibinfo {volume} {4}},\ \bibinfo {pages} {3817} (\bibinfo {year} {2010})}\BibitemShut {NoStop}%
\bibitem [{\citenamefont {Zhao}\ \emph {et~al.}(2018)\citenamefont {Zhao}, \citenamefont {Wu},\ and\ \citenamefont {Su}}]{Zhao:2018aa}%
  \BibitemOpen
  \bibfield  {author} {\bibinfo {author} {\bibfnamefont {M.}~\bibnamefont {Zhao}}, \bibinfo {author} {\bibfnamefont {W.}~\bibnamefont {Wu}},\ and\ \bibinfo {author} {\bibfnamefont {B.}~\bibnamefont {Su}},\ }\bibfield  {title} {\bibinfo {title} {ph-controlled drug release by diffusion through silica nanochannel membranes},\ }\href {https://doi.org/10.1021/acsami.8b12200} {\bibfield  {journal} {\bibinfo  {journal} {ACS Applied Materials \& Interfaces}\ }\textbf {\bibinfo {volume} {10}},\ \bibinfo {pages} {33986} (\bibinfo {year} {2018})}\BibitemShut {NoStop}%
\end{thebibliography}%

\end{document}